# Highly Integrated Organic-Inorganic Hybrid Architectures by Non-Covalent Exfoliation of Graphite and Assembly with Zinc Oxide Nanoparticles

Mario Marcia[a],[b], Chau Vinh[a], Christian Dolle[c], Gonzalo Abellán[a],[b], Jörg Schönamsgruber[a], Torsten Schunk[a], Benjamin Butz[c], Erdmann Spiecker[c], Frank Hauke[a],[b], Andreas Hirsch[a],[b],*

[a] M. Marcia, C. Vinh, Dr. G. Abellán, Dr. J. Schönamsgruber, Dr. T. Schunk, Dr. F. Hauke and Prof. Dr. A. Hirsch
Department of Chemistry and Pharmacy, Chair of Organic Chemistry II
Friedrich-Alexander-Universität Erlangen-Nürnberg
Henkestraße 42, 91054 Erlangen, Germany
E-mail: andreas.hirsch@fau.de

[b] M. Marcia, Dr. G. Abellán, Dr. F. Hauke and Prof. Dr. A. Hirsch
Joint Institute of Advanced Materials and Processes (ZMP)
Friedrich-Alexander-Universität Erlangen-Nürnberg
Dr.-Mack Str. 81, D-90762 Fürth, Germany

[c] C. Dolle, Dr. B. Butz and Prof. Dr. E. Spiecker
Institute of Micro- and Nanostructure Research
Friedrich-Alexander-Universität Erlangen-Nürnberg
Cauerstrasse 6, 91058 Erlangen, Germany

**Abstract:** Herein, we report an easy, straightforward, and versatile approach to build 0D/2D hybrid nanoparticle/graphene architectures by means of non-covalent chemistry and a modified Layer-by-Layer assembly. Three water soluble perylene diimides were employed to efficiently exfoliate pristine graphite into positively charged few- and multilayer graphene flakes. Further combination of these cationic building blocks with anionic zinc oxide nanoparticles led to the formation of tailor-made hybrid films *via* electrostatic and van der Waals interactions. These supramolecular hybrid nano-structures were thoroughly characterized by UV/Vis and Raman spectroscopy, AFM as well as electron microscopy, showing outstanding long-range homogeneity and high integrity in the centimetre-scale, uniform nanometric thickness between 60 – 100 nm and a close contact between the different building blocks. Due to their straightforward assembly. These architectures can be considered as promising candidates for numerous advanced applications especially in the field of energy storage and conversion.

## Introduction

Graphene is a high performance nanomaterial, both what fundamental physical properties and practical applications is concerned.[1] In order to fully exploit its promising potential, chemical functionalization, and integration into hybrid-architectures is of great importance. Recently, a couple of first proto-type functionalization protocols have been developed and examples for their implementation into new multilayer heterostructures have been accomplished.[2] Among others, the construction of nanoparticles-graphene (NP/G) hybrids has been proposed allowing the combination of the complementary properties of the different counterparts as well as the generation of intriguing synergistic effects.[3] In particular, the integration of graphene-based moieties with metal oxides such as zinc oxide-(ZnO)-NP is very appealing.[4] As a matter of fact, ZnO is a large-band-gap semiconductor. It is photocatalytically active under UV irradiation and at the same time highly biocompatible. Therefore, novel ZnO-NP/G hybrid nano-structures have been recently proposed with the purpose of applications in the field of energy storage and conversion[5] as well as in biosensors.[6] Furthermore, the chemistry of ZnO-NP is very well established in our group as we have developed several methods for their chemical functionalization in solution.[7] Up to date, the most common strategies for bottom-up synthesis of ZnO-NP/G hybrid structures, where G is generally reduced graphene oxide (rGO), rely on *in situ*[8] and *ex situ*[9] methods as well as on several chemical/electrochemical deposition approaches.[10] Additionally, the formation of hierarchical architectures of ZnO-NP and rGO has been reported lately.[11] However, the reduction of GO is never complete and this limits greatly the electronic mobility within the flakes. Moreover, the deposition density as well as the uniformity of the NP on GO are always strongly affected by the number of oxygen-containing functional groups and therefore by the quality of the GO material.[12] Therefore, and in order to preserve the sp$^2$-carbon network of pristine graphene layers, non-covalent exfoliation of graphite both in organic and aqueous media is an attractive choice.[2a] Along these lines our group has developed the chemistry of water-soluble perylene diimide (PDI)-based surfactants.[13] Their successful application as dispersing candidates has been reported recently for the individualization of carbon nanotubes[14] and exfoliation of graphite.[15] Unfortunately, the incorporation of unoxidized exfoliated graphene within NP/G hybrid materials is not trivial, due to its very limited solubility and processability.

To face this challenge, we herein present the facile and versatile formation of new graphene-based hierarchical architectures by means of non-covalent chemical functionalization of pristine graphite with tailor-made PDI surfactants followed by a modified Layer-by-Layer (LbL) assembly. Based on our easy, versatile and straightforward iterative dip-coating process, we were able to achieve functional multilayered architectures by assembling 2D-positively-charged-graphene-based building blocks with the 0D anionic ZnO-NP *via* electrostatic and van der Waals interactions. This conjugation leads to the formation of highly integrated organic-inorganic hybrid architectures exhibiting an intimate contact between their constituting building blocks. To the best of our knowledge, this is the first report concerning the preparation of tailor-made 0D/2D NP/G hybrid nano-structures through the implementation of a modified LbL-assembly procedure starting from pristine exfoliated graphite-based nanosheets. Furthermore, the use of non-covalent functionalization with cationic PDIs allowed us to precisely control the introduction of positive charges into the exfoliated carbon material without implying the disruption of the sp$^2$-carbon network.

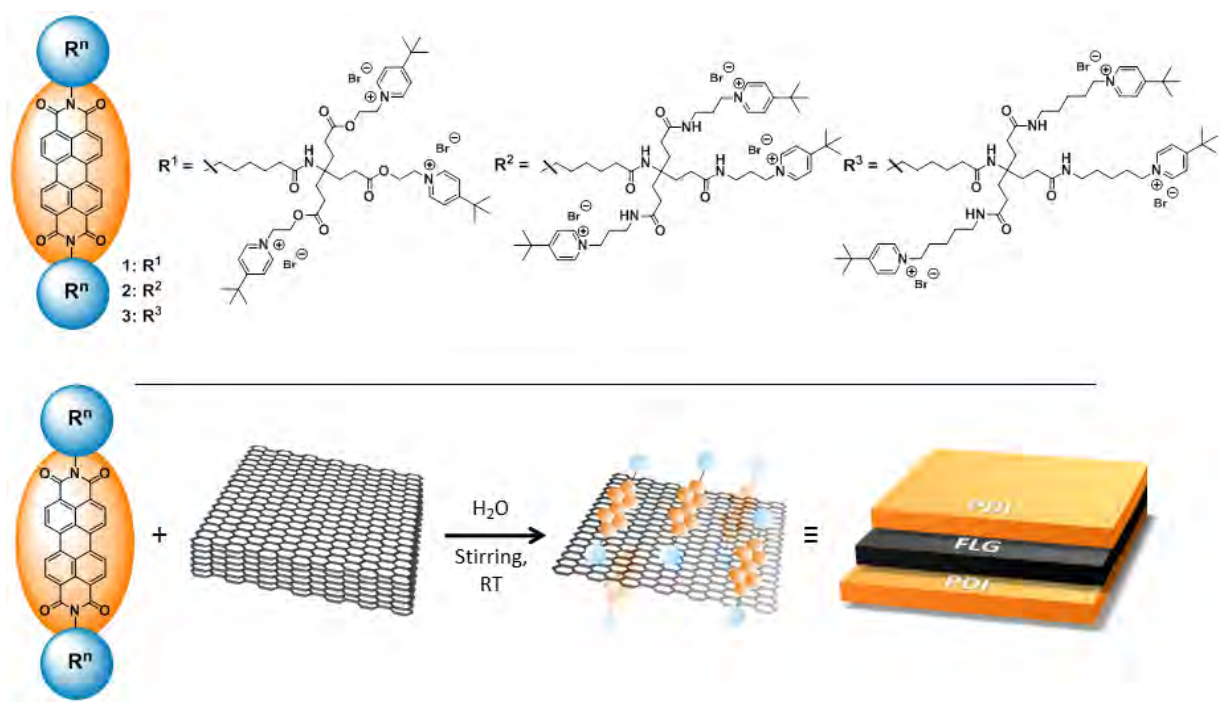

**Figure 1.** Top: Water soluble perylene diimide (PDI) based surfactants for the non-covalent functionalization of graphene. Bottom: Graphical representation for the formation of PDI-functionalized few layer graphene (FLG) by PDI-assisted exfoliation of graphite – no ultrasonic treatment is needed for the graphite delamination process.

**Results and Discussion**

Our approach is straightforward and based on ubiquitously available graphite. First of all, fresh dispersions of turbostratic graphite in bi-distilled water were prepared with the aid of cationic PDIs (**1**–**3**) by gentle magnetic stirring followed by centrifugation (Figure 1). Due to the intrinsic permanent stability of the positive charges of the terminating pyridinium units, PDIs (**1**–**3**) are readily soluble in pure water. This excludes the need to employ high ionic strength buffer solutions for the exfoliation experiments, as previously reported by Englert *et al.*,[15] in order to avoid contamination from salt deposits which could be detrimental for the iterative dip-coating approach.

The synthesis protocol of the novel ester-type dendritic PDI derivative **1** is given in the experimental section and in the SI (section 1), whereas the other water soluble, amide-type cationic pyridinium terminated PDIs (PDI **2** and **3**) were synthesized according to our previously published procedure.[13d]

As explained thoroughly in the SI (section 2) and as briefly highlighted in the experimental section of the manuscript, several experiments have been carried on to optimize the exfoliation process. Briefly, the ratio between the concentration of graphite and that of the PDI has been varied until the successful non-covalent functionalization was achieved for dispersions with maximized carbon uptake and absence of free-PDI molecules. By means of UV/Vis and fluorescence spectroscopy measurements, the strong energy/electronic interactions between the PDIs and the exfoliated carbon nanomaterial were investigated in solution (SI).

Remarkably, PDI derivative **1** proved to be the most effective agent towards the exfoliation of graphite in comparison to PDIs **2** and **3**. In particular, PDI **3** was found to be the least effective dispersing agent due to its increased solubility and reduced tendency for aggregation in comparison with derivatives **1** and **2**. As a matter of fact, PDI **3** possesses longer and more flexible spacer units in its periphery, which hinder a too close packing of the PDI cores due to electrostatic repulsions and translates into a lower tendency to aggregation in comparison to PDI derivatives **1** and **2**.[13d] As discussed in the SI and as reported by An *et al.*,[16] a more monomeric character of the exfoliation agent is generally linked to a lower exfoliation ability and therefore lower graphene content in the respective dispersions. Additionally, the fluorescence spectroscopy measurements in solution also indicated that the electronic communication for PDI **3** with the exfoliated carbon nanomaterial were less prominent with respect to those obtained for PDIs **1** and **2** (SI). Furthermore, upon exfoliation with PDI **1**, the final concentration of exfoliated carbon material in solution reached values up to 0.06 mg/mL (SI), which translates into an exfoliation capability of about 5 %. This value is higher than those reported recently by Zhang *et al.*,[17] concerning the exfoliation of graphite with naphthalene diimide (NDI)-based surfactants in water by ultrasonication. Our approach also provides the fundamental advantage that only mild magnetic stirring is employed which avoids an excessive disruption of the exfoliated graphitic flakes, as it might happen upon prolonged sonication.[18] After deposition by drop-casting of 1 µL of the dispersion on a Si/SiO$_2$ wafer, the exfoliated and PDI-functionalized carbon nanomaterial was investigated in the solid state by Raman characterization (Figure 2).

**Figure 2.** Mean Raman spectra for PDI/FLG hybrids after drop-casting of 1 μL of the freshly centrifuged dispersion on a Si/SiO$_2$ wafer and after drying: PDI **1**, blue trace; PDI **2**, red trace; PDI **3**, black trace. A) Laser excitation at λ$_{exc}$ = 532 nm; B) λ$_{exc}$ = 633 nm. *Raman peaks corresponding to the PDI moiety. In both A and B pictures, the depicted Raman spectra represent the mean spectra of 1680 recorded point spectra and they were normalized with respect to the intensity of the G-band.

For each sample, Raman maps containing 1680 individual spectra were collected and the mean Raman spectra were compared. The in-depth spectroscopic characterization of these hybrids revealed the presence of a strong electronic communication between the adsorbed PDI moieties and the exfoliated carbon material.

The Raman analysis of the different hybrids, measured with an excitation wavelength of 532 nm (Figure 2A), shows the coexistence of the PDI spectra and that of the exfoliated material in perfect agreement with the results of Kozhemyakina *et al.*[19] The observation of a minor contribution of a residual fluorescence background of the PDIs is also possible, although the electronic interaction results averagely in a significant fluorescence quenching. The non-covalent functionalization of the extended sp$^2$-carbon π-surface by means of the formation of PDI surface adsorption also resulted in a red-shift of the positions of the G- and 2D-band of the exfoliated material and led to a *p*-doping of this carbon allotrope.[20] In addition, the intensity and the full width at half maximum (FWHM) of the 2D-band is also drastically affected. As a direct consequence, the 2D-band appeared to be broadened in comparison to pristine un-functionalized graphene sheets and the respective 2D/G intensity ratio yielded values smaller than 1, as usually reported for highly non-covalent functionalized graphene.[21] In order to better account for the quality of the exfoliated graphene material and to avoid the interference of absorption/emission phenomena from the attached PDI molecules, samples were also characterized with a laser excitation of 633 nm (Figure 1B). In fact, as reported by Kozhemyakina *et al.*,[19] at this wavelength the absorption of the PDI core is greatly reduced and only a few of its residual Raman bands may be superimposed on the Raman profile of the exfoliated nanocarbon material. As depicted in Figure 2B, the 2D/G intensity ratio is smaller than 1 and the FWHM of the 2D-band is greater than 40 cm$^{-1}$, which clearly indicates that the non-covalent functionalization driven exfoliation process led to the formation of PDI-functionalized few and multilayer graphene (PDI/FLG and PDI/MLG) hybrid nano-structures.[22] Generally, the term FLG is attributed to aggregates consisting of less than 10 layers.[23] Nevertheless, the thickness of MLG flakes can vary between 10 and 100 layers.[24] Due to the fact, that the thickness of our final functionalized graphene material is comprised between 10 and 15 layers (averagely 12 layers from electron microscopy measurements, *vide infra*) we will define it as PDI/FLG in the following.

In order to acquaint for the stability of the PDI/FLG hybrids against washing in the solid state, the wafers prepared for the Raman characterization were washed by spin coating with water (1 mL) and with methanol (1 mL). After each washing step, the samples were dried and Raman characterization was performed ($\lambda_{exc}$ = 532 nm). As shown in the SI, the PDI/FLG hybrids resulted to be very stable against washing, as the characteristic Raman features of both PDI and FLG components are always present in the spectra after each washing step. This finding confirms the suitability of PDI/FLG hybrids as starting building blocks for implementation in the iterative coating-based assembly with ZnO-NP. The modified LbL assembly was performed by dip-coating a cleaned glass slide alternately into a solution of ZnO-NP in ethanol and into an aqueous dispersion of PDI (**1**–**3**)/FLG sheets. ZnO-NP of *ca.* 5 nm in diameter were synthesized *via* precipitation in ethanolic solution as previously reported by our group.[7c] The preparation protocol for the synthesis of the ZnO-NP is presented in the experimental section of the manuscript, while their characterization is given in the SI (section 3). In contrast to classical LbL approaches, no polymer layers were employed in our study in order to prevent a deterioration of the intimate contact between the different building blocks.

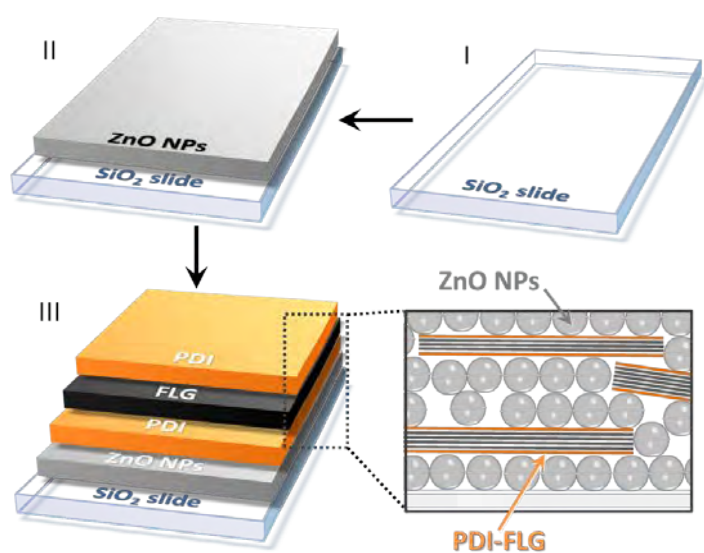

**Figure 3.** Graphical representation of the fabrication of the [ZnO-NP/PDI/FLG]$_n$ hybrid nano-structures by means of an iterative dip-coating approach: **I**) Cleaned glass slide; **II**) deposition of a layer of ZnO-NP; **III**) formation of the [ZnO-NP/PDI/FLG]$_n$ hybrid structure. The highlighted section depicts a schematic representation of the intricated assembly of the ZnO-NP and the PDI/FLG building blocks obtained in the course of the assembly approach – structure of the hybrid clarified by HRTEM investigations.

As depicted in Figure 3, ZnO-NP were first deposited on a cleaned glass slide (**I**) which was subsequently rinsed with ethanol and dried with compressed air. Afterwards, the slide coated with the ZnO-NP (**II**) was immersed in the PDI/FLG aqueous dispersion, then rinsed with bi-distilled water and dried to obtain the first hybrid nano-structure ([ZnO-NP/PDI/FLG]$_1$, **III**). After four repetitions of this procedure, a hybrid graphene-based architecture could be obtained *via* electrostatic and van der Waals interactions between the anionic charged ZnO-NPs and the positively charged groups of the PDI moieties, [ZnO-NP/PDI/FLG]$_4$.

The growth was stopped after 4 repetitions in order to prepare ultra-thin films which could be successfully investigated by electron microscopy. Indeed, the strong π-π stacking interactions between the PDI core and the basal plane of the extended π-carbon allotrope allowed the efficient incorporation of functionalized FLG sheets into the hybrid nanomaterial assembly. At this point, it is worth to mention that upon integration of the PDI/FLG sheets by successive coating, further re-aggregation by π-π stacking interactions[25] might take place and lead to the presence of highly dispersed and functionalized graphitic flakes with heterogeneous thickness within the [ZnO-NP/PDI/FLG]$_4$ hybrid nano-architectures, as observed by electron microscopy measurements (*vide infra*). The assembly of the [ZnO-NP/PDI/FLG]$_n$ nano-structures was monitored by UV/Vis spectroscopy. By plotting the absorption at 338 nm (ZnO-NP absorption) and at 498 nm (PDI absorption) against the number of repetitions we obtained a linear correlation. As depicted in Figure 4, the experimental values for the absorption collected for PDI **1** could be fitted linearly at both wavelengths. Such good linear correlations are generally indicative for the successful LbL-deposition and point towards the formation of highly hierarchical hybrid architectures. Similar results were obtained for PDI **2** and **3** (SI, section 4).

**Figure 4.** Experimental values and linear fits of the UV/Vis absorption of ZnO-NP/PDI/FLG assemblies at 338 nm (squares) and at 498 nm (dots), against the number of repetitions (PDI = **1**).

However, as previously depicted, the interaction of PDI **3** with the FLG sheets was less prominent in comparison with PDI **1** and **2**. For this reason, the in-depth sample characterization of the [ZnO-NP/PDI/FLG]$_4$ nano-architectures was only carried out with hybrid assemblies obtained with PDI **1** and **2** as exfoliating and dispersing agents.

In order to obtain more information about the intimate structure within the hybrid nano-structures of the [ZnO-NP/PDI/FLG]$_4$ with PDI **1** or **2**, Raman spectroscopy, atomic force microscopy (AFM), scanning electron microscopy (SEM), and transmission electron microscopy (TEM) measurements have been carried out. Raman characterization was performed systematically in order to analyze the quality and homogeneity of the deposited material after the iterative dip-coating process. Raman maps were recorded which showed that the surface of the assembled hybrid nano-structures is highly homogeneous for, at least, areas of about 40 µm$^2$. Indeed, Figure 5A shows a 19 x 20 µm$^2$ mapping highlighting the ratio between the PDI peak located at 1383 cm$^{-1}$ and the G-band from FLG ($I_{PDI}/I_G$)e. The mean spectra for the [ZnO-NP/PDI/FLG]$_4$ assemblies with PDI **1** or **2** are presented in Figure 5B. In both spectra, the typical features of the PDI/FLG hybrids are clearly discernible, whereas the position of the 2D-band appears to be red-shifted of ca. 20 cm$^{-1}$ in comparison to Figure 2A.

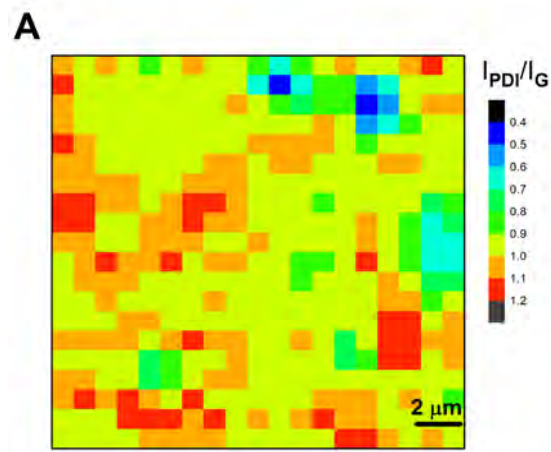

**Figure 5.** A) Raman intensity map of the $I_{PDI}/I_G$ ratio, PDI = **1**, 19 x 20 µm$^2$. $I_{PDI}$ is referred to the intensity of the PDI peak located at 1383 cm$^{-1}$. B) Mean Raman spectra for the [ZnO-NP/PDI/FLG]$_4$ hybrid nano-architectures: PDI **1**, blue trace; PDI **2**, red trace; $\lambda_{exc}$ = 532 nm. *Raman peaks corresponding to the PDI moiety. The Raman spectra are the mean value of 36 recorded spectra (measured area ca. 40 µm$^2$) and they were normalized for the intensity of the G-band.

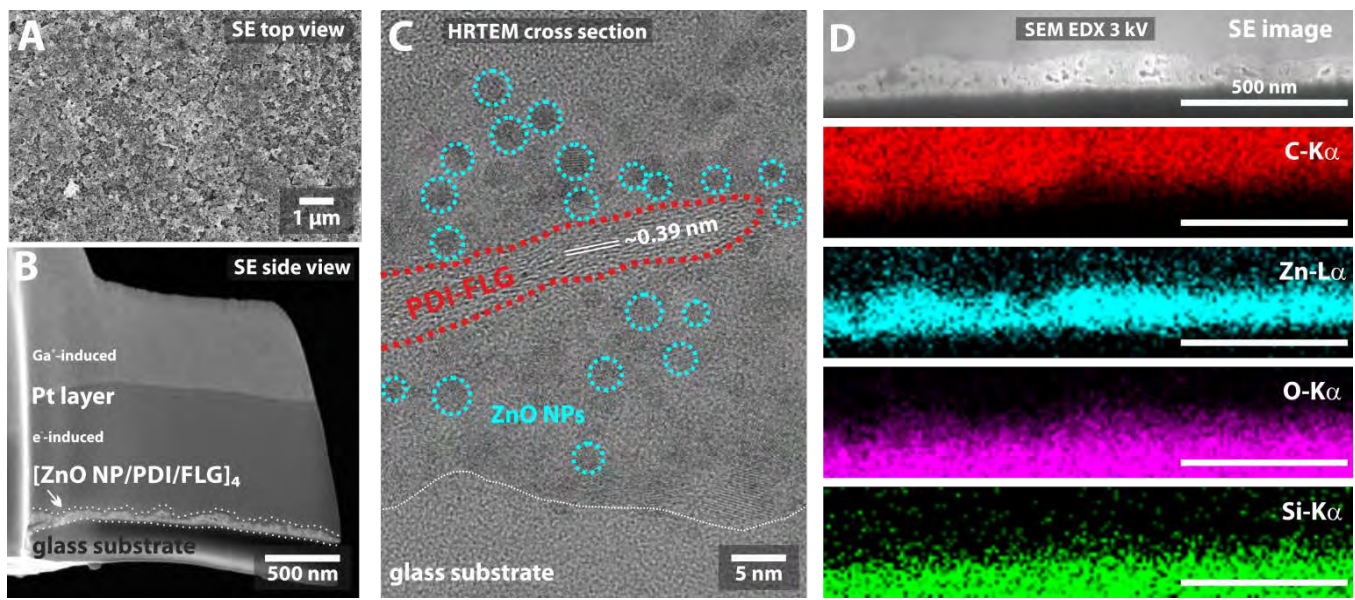

**Figure 6.** A) Low-magnification SEM image showing the top view of the film. B) SE image of the final cross-section lamella prepared by focused ion beam. C) HRTEM of mixed film, FLG, and several ZnO-NP are highlighted. D) SE image and EDX mappings at 3 kV in the SEM showing the spatial distribution of C, Zn, O, and Si in the investigated area.

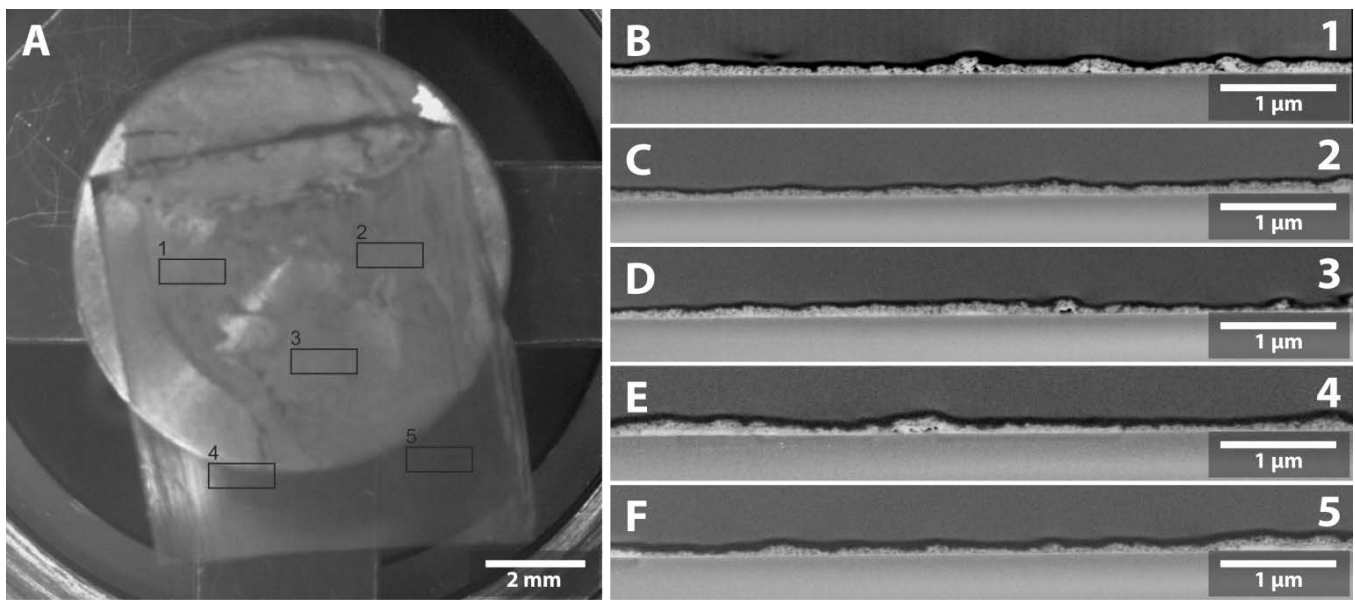

**Figure 7.** Cross sectional analysis of [ZnO-NP/PDI/FLG]$_4$ with PDI **1** on different positions of the sample. Photo of the sample geometry highlighting the measured cross sections (A), position 1 (B), position 2 (C), position 3 (D), position 4 (E), position 5 (F), showing a remarkable homogeneity of layer thickness.

To gain a deeper insight in the hybrid nano-structures, AFM characterization was performed (SI), which confirmed the homogeneity of the [ZnO-NP/PDI/FLG]$_4$ hybrid films (within 10 µm$^2$). This represents a good component distribution and sample homogeneity as the PDI/FLG flakes have a moderate dimension within the hybrid. It is also worth to take into account that films grown by LbL-like processes might lead the introduction of imperfections (such as holes) in the material, as already reported for similar oligoelectrolyte multilayer films by Rosenlehner et al.[26] Furthermore, it was determined that the surface of the hybrids was smoother in presence of FLG sheets as the average root mean square roughness was found to be about 6.5, significantly lower than that measured without FLG sheets, *i.e.* 10.5 (SI). In order to further analyze the morphology and distribution of the different building blocks within the film, we used a combination of FIB/SEM and HRTEM, employing a FEI Helios NanoLab660 and a Philips CM300 as well as a FEI Titan Themis³ 300 microscope. The top-view (imaging in SEM with secondary electrons, Figure 6A) of the coated glass slide reveals a homogeneous covering of the whole surface area with a certain roughness (see also AFM characterization in the SI). To evaluate the distribution of the compounds within the layered system a lift-out lamella was prepared and investigated subsequently by TEM at 300 kV (CM300) or 200kV (FEI Titan Themis³ 300) accelerating voltage. To enhance the conductivity of the films and to facilitate the lamella preparation two carbon threads (*ca*. 14 nm thick) were evaporated onto the sample in order to minimize charging during the ion-milling. For protection of the sensitive layers a bar of Pt precursor was deposited *via* electron beam deposition, followed by ion beam deposition and two trenches were milled using Ga$^+$ ions at 30 kV accelerating voltage (Figure 6B). To check the thickness of the film several trenches were milled at different positions on the glass slide confirming an outstanding homogeneous thickness between 60–100 nm (Figure 7 and SI). A lamella of the hybrid nano-structure of [ZnO-NP/PDI/FLG]$_4$ with PDI **1** was prepared, attached to an Omniprobe-grid and thinned to a thickness of 80 nm for the HRTEM investigation (SI). A spatially resolved EDXS mapping was conducted using the SEM at 3 keV beam energy (Figure 6D), confirming the Zn-containing nanoparticles of ZnO on the SiO$_2$ substrate. The HRTEM investigation showed a distribution of ZnO-NP with a size of approximately 5 nm with embedded graphitic flakes, supporting the model picture of the [ZnO-NP/PDI/FLG]$_4$ hybrid nano-structure presented in Figure 3, in clear contrast with the idealized picture usually associated to films grown by classical LbL methods. Further analysis revealed the presence of an intricate structure consisting of a flake with a thickness of *ca*. 5 nm (~13 layers) and interlayer distance of ~0.39 nm (Figure 6C) completely surrounded by a ZnO-NP matrix (see also SI, section 4, for additional measurements).

The crystalline nature of the ZnO-NP was confirmed by the HRTEM investigation, corroborating the results of the EDXS. The 5 nm size of the NP is also in good agreement with the acquired DLS data presented in the SI. Moreover, this study reveals that the assembly of the 0D NP and the non-covalent functionalized 2D FLG leads to the formation of highly homogeneous centimeter-scale thin films consisting of entangled structures with an improved interfacial contact between the different constituents in the nanometric range – a matter of utmost importance to develop new functional materials.

**Conclusions**

In summary, by the exfoliation of graphite and the attachment of tailor-made cationic PDI surfactants to FLG flakes, positively charged functionalized 2D graphene-based hybrids are formed. These building blocks can easily be conjugated with anionic 0D ZnO-NP by means of a straightforward and versatile iterative dip-coating process. The full characterization of these hybrid supramolecular architectures revealed the formation of ultra-thin films with outstanding thickness, centimeter-scale homogeneity, and with an entangled structure exhibiting a close contact between the organic and inorganic building blocks. Due to their low cost and high processability, the reported [ZnO-NP/PDI/FLG]$_n$ hybrid nano-architectures exhibit a interesting potential for application in the field of energy storage and conversion. Further work along this line is currently underway in our laboratories.

**Experimental Section**

**Materials**

Commercially available reagents were purchased from Sigma Aldrich and, if not otherwise noted, used without any further purification. All analytical-reagent grade solvents were purified by distillation. Turbostratic graphite powder (serial nr. 2449.363) and sulfuric acid (98%) were purchased from VWR International Inc.; bi-distilled water from Carl Roth Inc.

**Synthesis**

**Synthesis and charcterization of PDI 1:** 4-(tert-Butyl)-1-(2-hydroxyethyl)pyridinium bromide was synthesized according to Kuhnen-Clausen *et al.*[27] while the precursor PDI, whose structure is reported in Figure S1 in the SI, according to Schmidt *et al.*[13a]

In a 50 mL round bottom flask 0.25 g (0.23 mmol) of the precursor PDI and 0.38 g (2.79 mmol) HOBt were dissolved in 20 mL dry DMF. After cooling to 0 °C, 0.57 g (2.79 mmol) of DCC were added and the solution was stirred for 15 min. 0.72 g (2.79 mmol) of 4-(tert-butyl)-1-(2-hydroxyethyl)pyridinium bromide were subsequently added and the solution was furthermore stirred for three days. Afterwards the formed precipitate of DCU was filtrated and the solvent removed under reduced pressure. The residue was dissolved in a small amount of MeOH and precipitated with acetone for several times. Yield: 0.43 g (0.17 mmol, 74 %).

**$^1$H-NMR:** (400 MHz, CD$_3$OD, rt): δ = 1.33 (m, 4 H, CH$_2$), 1.49 (s, 54 H, CH$_3$), 1.60 (m, 4 H, CH$_2$), 1.98 (m, 12 H, CH$_2$), 2.39 (m, 4 H, CH$_2$), 2.57 (m, 12 H, CH$_2$), 2.87 (m, 4 H, CH$_2$), 3.66 (m, 4 H, CH$_2$), 4.66 (m, 12 H, CH$_2$), 5.05 (m, 12 H, CH$_2$), 7.34 (m, 8 H, Ar-H), 8.28 (d, J = 6.4 Hz, 12 H, Ar-H), 9.10 (d, J = 5.92 Hz, 12 H, Ar-H) ppm. **$^{13}$C-NMR:** (100 MHz, CD$_3$OD, rt): δ = 25.54, 27.86, 28.15 (6 C, CH$_2$), 30.04 (6 C, $_{CH2}$), 30.34 (6 C, CH$_2$), 30.60 (18 C, CH$_3$), 34.28 (2 C, CH$_2$), 38.03 (6 C, C, q), 39.56 (2 C, CH$_2$), 60.93 (2 C, C, t), 64.35 (6 C, CH$_2$), 68.87 (6 C,

CH$_2$), 122.60, 124.64, 126.45, 127.04, 128.40, 131.24, 133.78, 146.46 (44 C, Ar-C), 163.81 (4 C, C=O), 173.58 (6 C, C=O), 174.23 (6 C, Ar-C), 174.77 (2 C, C=O) ppm. **MS (MALDI):** m/z = 1931 [M-2 branches-3 Br]$^+$. **EA:** C$_{122}$H$_{156}$Br$_6$N$_{10}$O$_{18}$.H$_2$O: calcd. (%): C 57.51, H 6.25, N 5.50; found: C 57.12, H 6.11, N 5.72. **IR (ATR):** $\tilde{v}$ = 3324, 2932, 2852, 1692, 1654, 1625, 1577, 1346, 1244, 811 cm$^{-1}$. **UV/Vis** (H$_2$O): λ$_{max}$ = 320, 475, 498, 540 nm.

**Synthesis of PDI 2 and 3:** The synthesis of these two derivatives was performed according to Schönamsgruber *et al.*[13d]

**Preparation of the PDI-FLG hybrid:** Fresh solutions of PDIs **1–3** were prepared in bi-distilled water and used to disperse turbostratic graphite by mild magnetic stirring (1400 rpm) at room temperature. Typically, 10 mL-snap cap vials were employed. By carefully controlling the ratio between the PDI and graphite, the best parameters were successfully defined where all the PDI molecules are attached to the exfoliated material in solution and no free PDI is left (c$_{G, I}$ = 1.2 mg/mL; c$_{PDI}$ = 10$^{-5}$ M). The exfoliation was further optimized by time–dependent exfoliation experiments. Finally, the stability of the PDI–FLG hybrid against washing was proven by Raman spectroscopy. For detailed explanations, please refer to the SI (section 2).

**Synthesis of the ZnO-NP:** ZnO-NP were synthesized according to a procedure previously reported by our group.[7c] Namely, 550 mg zinc acetate dehydrate was dissolved in 25 mL boiling ethanol and cooled down to 38 °C. Then, 125 mg LiOH was dissolved in 25 mL boiling ethanol and cooled down to 38 °C. Afterwards, the LiOH solution was added dropwise to the zinc acetate solution under vigorous stirring. After 20 h, the ZnO-NP were obtained with a size of *ca.* 5.6 nm and a zeta potential value of -7.8 mV, which is in accordance to that reported by Segets *et al.*[7f] For UV/Vis and DLS data, please refer to the SI (section 3).

**Preparation of the [ZnO-NP/PDI/FLG]$_4$ nano-architectures by an iterative dip-coating process:** First of all, the glass slides were cleaned in a solution of sulfuric acid (96%) and hydrogen peroxide (2:1) for 30 minutes. Afterwards, they were rinsed with distilled water until neutrality and then dried with compressed air. For the preparation of the assemblies, a solution of ZnO-NP (1 mg/mL) in EtOH and a fresh centrifuged dispersion of PDI/FLG in bi-distilled water were employed. The immersion time was chosen of 20 minutes, adapted to previous reports from our group.[26, 28] ZnO-NP were first deposited on a cleaned glass slide which was subsequently rinsed with ethanol and dried with compressed air. Afterwards, the slide coated with the ZnO-NP was immersed in the PDI/FLG aqueous dispersion, then rinsed with bi-distilled water and dried to obtain the hybrid nano-structure [ZnO-NP/PDI/FLG]$_1$. After four repetitions of this procedure, a hybrid graphene-based nano-architecture could be obtained, [ZnO-NP/PDI/FLG]$_4$. For additional UV/Vis spectroscopy, AFM as well as further electron microscopy characterization please refer to the SI (section 4).

**Characterization**

**$^1$H- and $^{13}$C-NMR spectroscopy** were performed on Jeol JNM EX 400 and JEOL JNM GX 400 ($^1$H: 400 MHz and $^{13}$C: 100.5 MHz), Bruker Avance 300 ($^1$H: 300 MHz and $^{13}$C: 75 MHz). The resonance multiplicities are indicated as s (singlet), d (doublet), t (triplet), q (quartet), m (multiplet) and br (broad signal). The raw data sets were processed using the software ACD/Labs 10.0 and MestReNova Lite 5.2.5 from Mestrelab Research S.L. **Mass spectrometry** was conducted on a Shimadzu AXIMA Confidence, MALDI-ToF mass spectrometer, Nitrogen UV-LASER (50 Hz, 337 nm). Matrix: 2,5-dihydroxybenzoic acid (DHB). **IR spectroscopy** was performed on a Bruker Vector 22 with an ATR RFS 100/S detector and the compounds were measured by coating the pure solid on a diamond crystal. **Elemental analysis** was performed on an EA 1110 CHNS from CE Instruments. **Centrifugation** of the PDI-FLG dispersions was carried out in a VWR 1814 Microcentrifuge at room temperature. All samples were centrifuged 30 min at 1000 rpm prior to spectroscopic characterization. **UV/Vis spectra** were recorded on a Perkin Elmer Lambda 1050 spectrometer (integration time 0.24 s; slit width 2 nm). **Fluorescence spectra** of the dispersions were measured with a Horiba Scientific Fluorolog-3 Spectrometer with a PMT detector (integration time 0.24 s; slit width 2 nm). **Raman maps** were recorded with a Horiba Jobin Yvon Aramis confocal Raman spectrometer equipped with a microscope (100x objective used) and an automated *XYZ*-table, using laser excitations of 532 and 633 nm. The PDI/FLG dispersions were analyzed by drop-casting 1 μL of the dispersion onto a Si/SiO$_2$ (300 nm) wafer and drying (30 min at 75 °C under vacuum), while the [ZnO-NP/PDI/FLG]$_4$ samples were studied as prepared on the glass slides. The measurements were performed with an integration time of 1 s, filter D1 and 600 gr/mm gratings. For the analysis of the [ZnO-NP/PDI/FLG]$_4$ samples the hole and slit were kept half opened. **DLS and Zetapotential** were recorded on a Zetasizer Nano series ZEN3600 (Malvern Instrument Ltd. UK) dynamic light scattering analyser. **AFM** analyses were performed on the [ZnO-NP/PDI/FLG]$_4$ samples as prepared on the glass slides in a Veeco di 300, nanoscope IIIa equipment. **Electron Microscopy:** A dual-beam instrument (FEI Helios NanoLab 660) Focused Ion Beam (SEM/FIB) has been used for imaging and site-specific preparation of cross-sectional slices. After the *ex-situ* evaporation of two carbon threads (ca. 14 nm) a suitable area for cross-sectioning was selected. Afterwards, a layer of Pt (> 1 μm) was deposited in the SEM by electron beam, followed by deposition with the ion beam, in order to protect the area of interest during the lamella preparation. Two trenches were milled by FIB to obtain a lamella, which was transferred to an Omniprobe grid using an Easy-lift NanoManipulator and thinned with the ion beam. The final thickness of the lamella was kept below 100 nm, to allow its characterization by HRTEM. Transmission electron microscopy (TEM) was performed using a CM300 UltraTWIN (Philips, Netherlands), equipped with a LaB6 filament and a nominal point resolution of 1.7 Å at the Scherzer defocus. The microscope was

operated at an acceleration voltage of 300 kV. TEM images were recorded with a charged coupled device camera (TVIPS, Germany), which has an image size of 2048 × 2048 pixels. Selected analyses were performed on an aberration corrected FEI Titan Themis³ 300 TEM with a high brightness field emission gun (X-FEG) operated at 200kV. EDXS measurements were carried in a FEI Helios NanoLab 660 using a large area Oxford SDD-detector (X-Max$^N$). The beam energy was 3 keV allowing to probe the specific X-Ray signals of C-K□ (270 eV), O-K□ (520 eV), Zn-L□ (1020 eV) and Si-K□ (1740 eV). The integrated intensity of the respective signal was then used to construct spatially resolved maps showing the elemental distribution.

## Acknowledgements

The authors thank the Deutsche Forschungsgemeinschaft (DFG – SFB 953 "Synthetic Carbon Allotropes" – projects A1 and Z2), the Cluster of Excellence Engineering of Advanced Materials (EAM) and the Interdisciplinary Center for Molecular Materials (ICMM) for the financial support. G.A. thanks the EU for a Marie Curie Fellowship (FP7/2013-IEF-627386). In addition, the research leading to these results has received partial funding from the European Union Seventh Frame-work Program under Grant Agreement No. 604391 Graphene Flagship. Finally, the authors thank Sebastian Etschel, from the groups of Prof. M. Halik and Prof. R. Tykwinski, for measuring AFM of the ZnO/PDI/FLG hybrids (SI).

Supporting Information

# Highly Integrated Organic-Inorganic Hybrid Architectures by Non-Covalent Exfoliation of Graphite and Assembly with Zinc Oxide Nanoparticles

Mario Marcia[a],[b], Chau Vinh[a], Christian Dolle[c], Gonzalo Abellán[a],[b], Jörg Schönamsgruber[a], Torsten Schunk[a], Benjamin Butz[c], Erdmann Spiecker[c], Frank Hauke[a],[b], Andreas Hirsch[a],[b],*

# 1. Synthesis of PDI

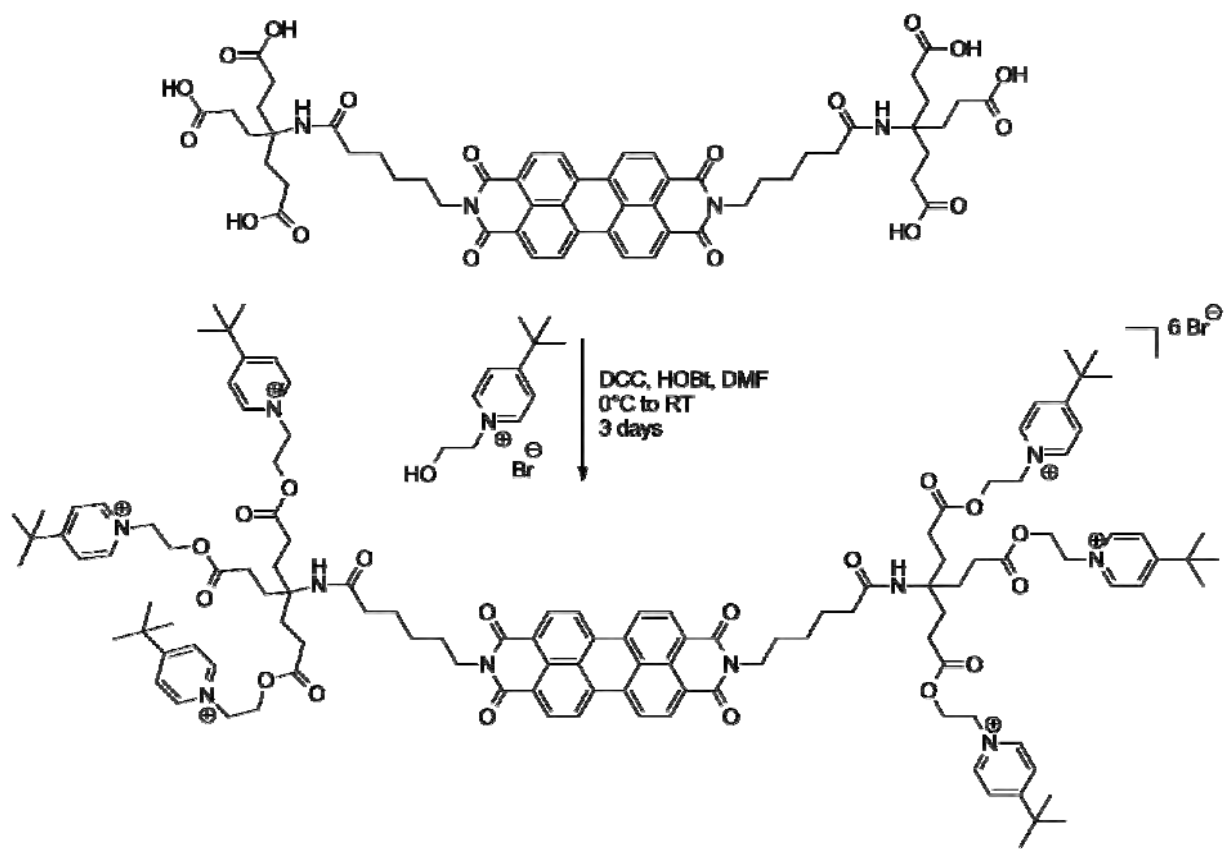

**Figure S1.** Scheme for the synthesis of the novel ester type PDI derivative **1**.

## 2. Synthesis and Characterization of the PDI/FLG Hybrids

### 2.1 Variation of the Concentration of the Graphite

Fresh solutions of PDIs **1**–**3** (c = $10^{-5}$ M) were prepared in bi-distilled water and used to disperse different amounts of turbostratic graphite (c = 0.12, 1.2 and 12 mg/mL) by mild magnetic stirring for 48 h, at room temperature. Afterwards, the dispersions were let stand still for ≈ 20 h to allow the sedimentation of the un-stabilized graphite species and then centrifuged. The characterization of the supernatants by absorption spectroscopy shows clearly that efficient exfoliation of graphite could be performed only in a very well defined range of concentrations (Figure S2).

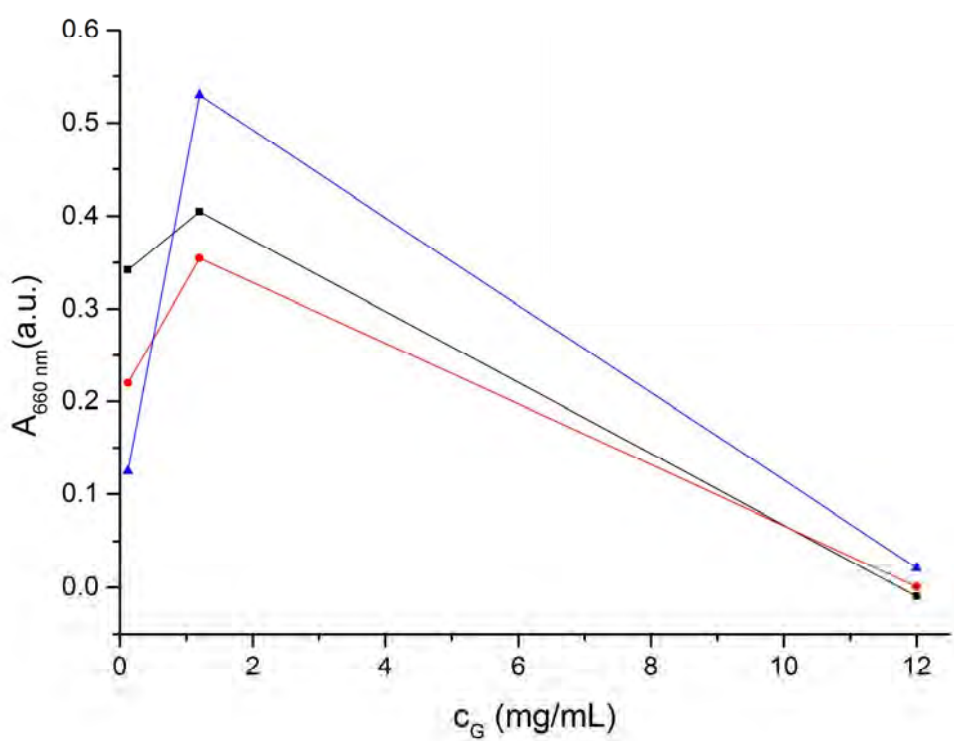

**Figure S2.** Relation between the absorption at 660 nm after centrifugation ($A_{660\ nm}$) and the initial concentration of graphite ($c_G$) for the dispersion of turbostratic graphite by means of different cationic PDIs in pure water (stirring time 48 h): PDI **1**, blue trace; PDI **2**, red trace; PDI **3**, black trace.

In the following, a detailed analysis of the dispersion experiments by employing PDI derivative **1** will be presented due to the fact that this derivative turned out to be the most effective for the exfoliation of graphite with subsequent non-covalent functionalization of the FLG sheets in solution. Nevertheless, the same considerations can be applied to PDI **2** and to some extents also to derivative **3**, as it will be explained later (see also the main text).

The absorption profiles of the respective **1**/FLG dispersions, upon variation of the graphite content (c = 0.12, 1.2 and 12 mg/mL) by keeping constant that of the PDI (c = $10^{-5}$ M), are plotted in Figure S3. On the one hand, when the amount of the graphitic starting material is too high, in comparison to that of the PDI (dotted line), the dispersions are not stable and the graphitic nanomaterial was almost completely precipitated after centrifugation. On the other hand, when the amount of graphite is too low in comparison to the concentration of PDI (dashed line), exfoliation and functionalization occur but a significant amount of free, unattached PDI molecules are present in solution (strong absorption around 450 – 600 nm). Therefore, a threshold ratio between the PDI surfactant and the graphite must be reached to allow a successful exfoliation and functionalization (straight line).

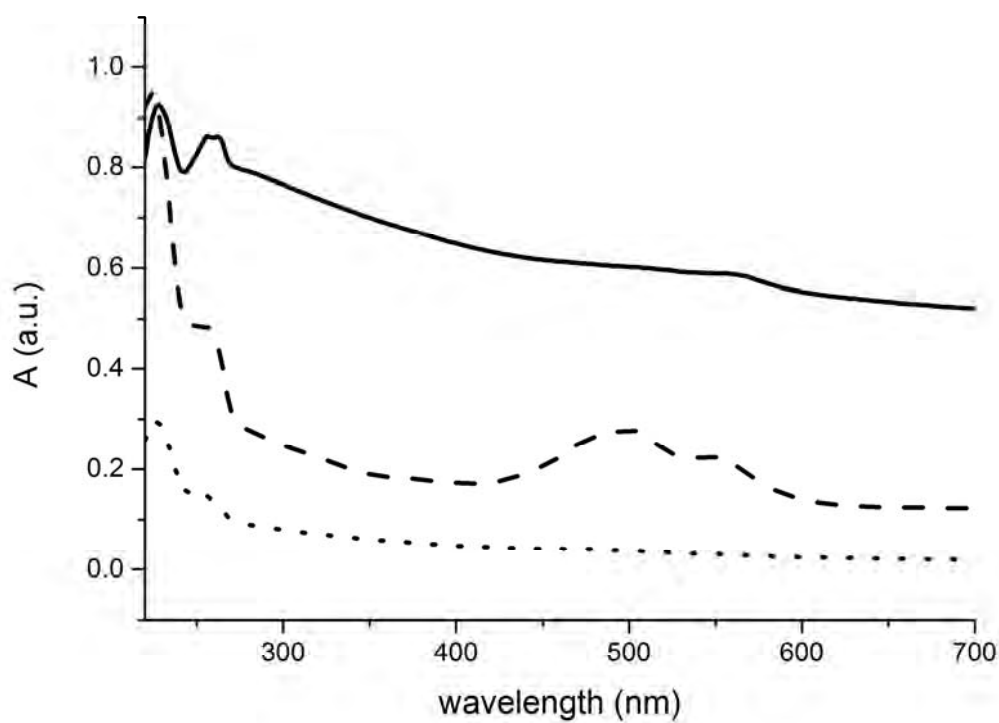

**Figure S3.** UV/Vis absorption spectra of **1**/FLG dispersions in pure water by varying the initial concentration of graphite ($c_G$ = 12 mg/mL, dotted line; $c_G$ = 0.12 mg/mL, dashed line; $c_G$ = 1.2 mg/mL, solid line), with PDI **1** as stabilizer ($c_1$ was kept constant at $10^{-5}$ M).

The presence of free PDI molecules also hinders the precise determination of the attachment of the PDIs to the exfoliated graphitic substrate and therefore the efficient non-covalent functionalization in solution. As a matter of fact, their dominant absorption masks the quenched red-shifted profiles of the PDI/FLG hybrids in solution also already shown in Figure S2. Moreover, free PDI molecules contribute to a greater fluorescence of the dispersions and thus to a more difficult determination of the true PDI/FLG interaction (Figure S4).

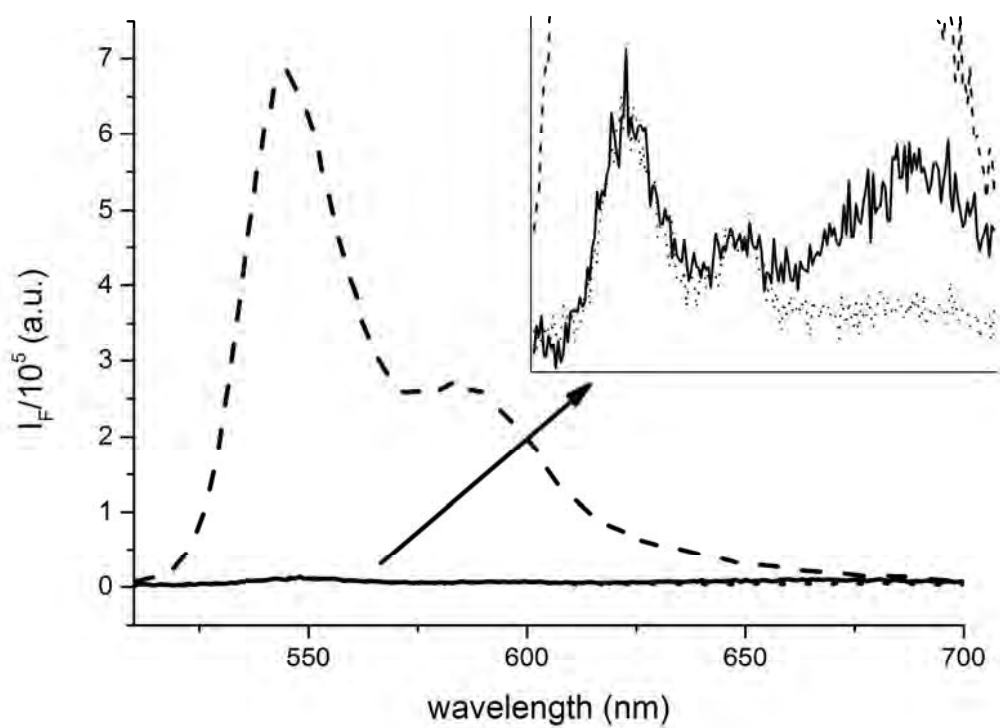

**Figure S4.** Fluorescence emission spectra of **1**/FLG dispersions in pure water by varying the initial concentration of graphite ($c_G$ = 12 mg/mL, dotted line; $c_G$ = 0.12 mg/mL, dashed line; $c_G$ = 1.2 mg/mL, straight line), with PDI **1** as stabilizer ($c_1$ was kept constant at $10^{-5}$ M).

## 2.2 Variation of the Concentration of the PDIs

The existence of a preferable concentration range to allow a successful dispersion and non-covalent functionalization of the graphitic starting material was also confirmed by additional exfoliation experiments. Different dispersions were prepared by varying the concentration of PDIs **1**–**3** in pure aqueous solution (c = $10^{-6}$, $10^{-5}$ and $10^{-4}$ M) and keeping constant the amount of turbostratic graphite (c = 1.2 mg/mL). As shown in Figure S5, an increase in the concentration of the PDI results in a higher carbon uptake in solution.

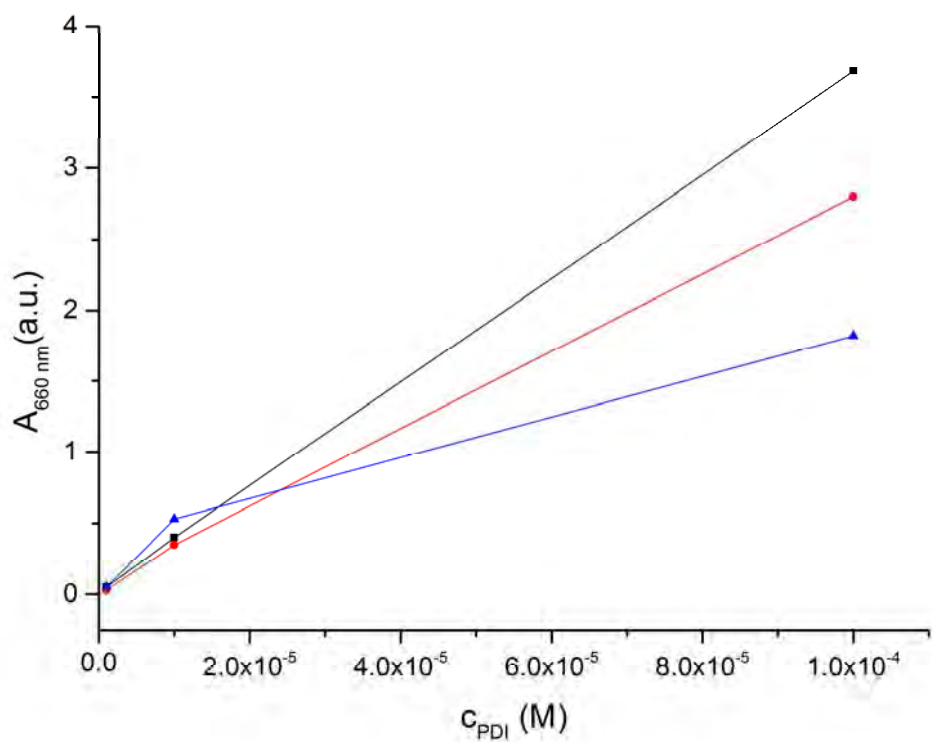

**Figure S5.** Relation between the absorption at 660 nm after centrifugation ($A_{660\ nm}$) and the concentration of the PDI surfactants ($c_{PDI}$) for the dispersion of turbostratic graphite by means of different cationic PDIs in pure water (stirring time 48 h): PDI **1**, blue trace; PDI **2**, red trace; PDI **3**, black trace.

An increase of the carbon uptake in solution, however, does not directly correlate with a successful interaction of the PDI with graphene. As a matter of fact, a too low concentration of the PDI in comparison to that of graphite determines an unsuccessful dispersing capability of the surfactant whereas a too high PDI/FLG ratio implies the presence of a large amount of free PDI molecules in solution. Therefore, high carbon uptakes, which indicate a high amount of carbon species in solution as a result of good exfoliation process, are not always accompanied by a successful non-covalent functionalization. The combination of both good exfoliation and functionalization occurs only when the PDI/FLG ratio reaches a proper threshold value. This concept can be visualized by comparing the absorption (Figure S6) and emission (Figure S7) spectra of **1**/FLG dispersions where this time the concentration of the PDIs has been varied.

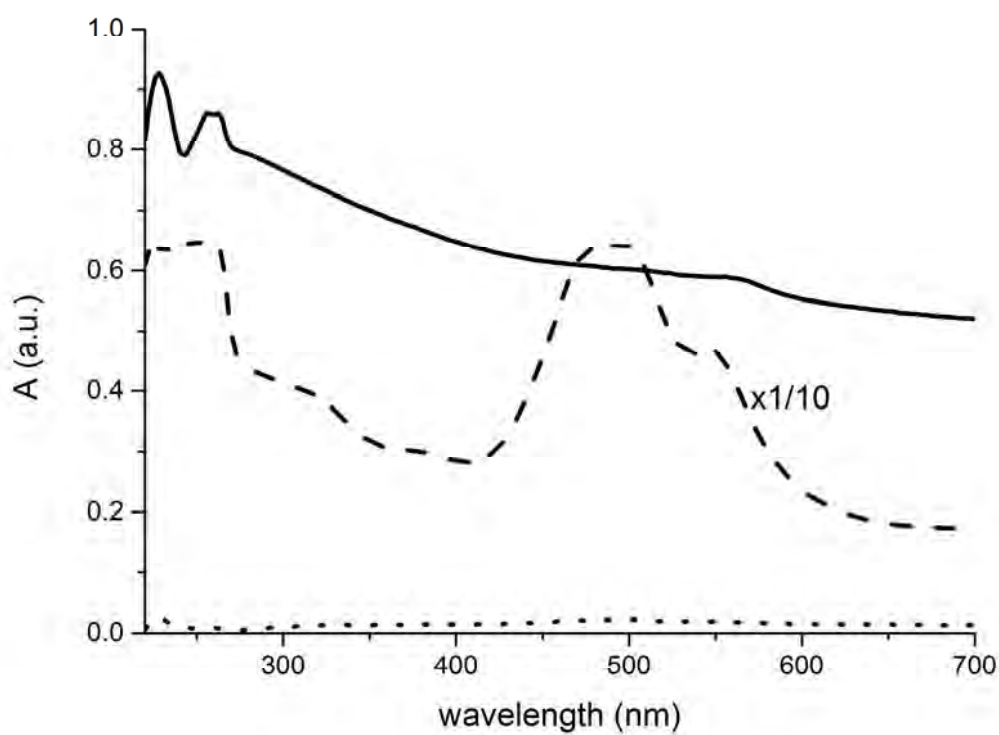

**Figure S6.** UV/Vis absorption spectra of **1**/FLG dispersions in pure water by varying the concentration of the PDI surfactant **1** ($c_1 = 10^{-6}$ M, dotted line; $c_1 = 10^{-4}$ M, dashed line; $c_1 = 10^{-5}$, solid line) and keeping the initial concentration of graphite constant at 1.2 mg/mL.

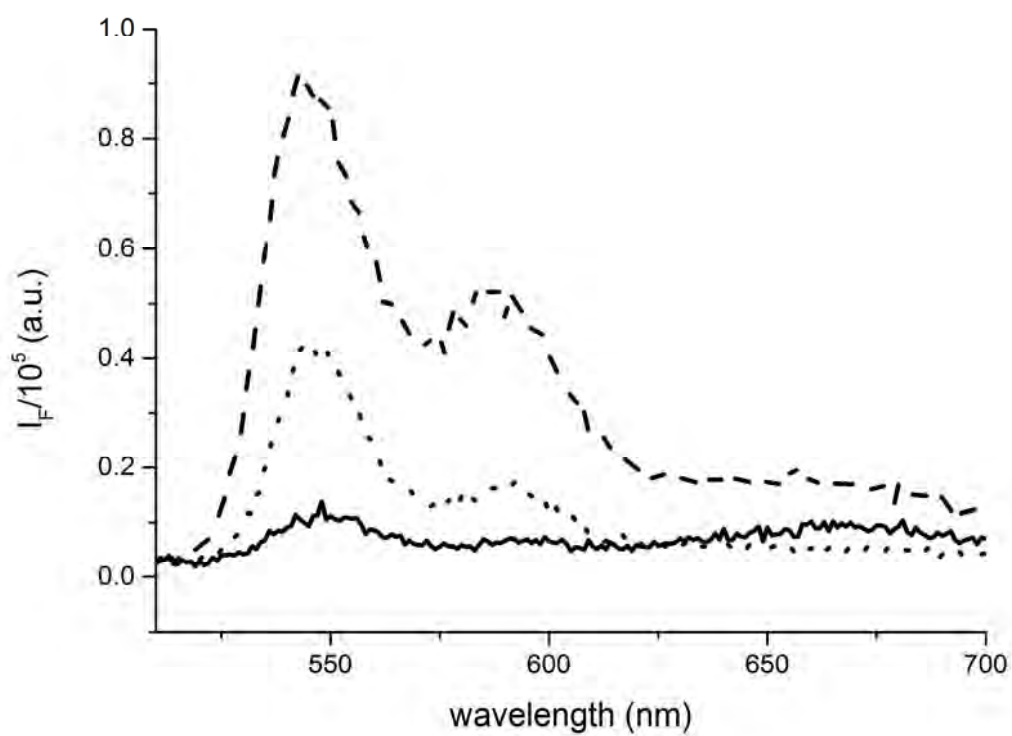

**Figure S7.** Fluorescence emission spectra of **1**/FLG dispersions in pure water by varying the concentration of the PDI surfactant **1** ($c_1 = 10^{-6}$ M, dotted line; $c_1 = 10^{-4}$ M, dashed line; $c_1 = 10^{-5}$, straight line) and keeping the initial concentration of graphite constant at 1.2 mg/mL; $\lambda_{exc} = 498$ nm.

## 2.3 Summary of the UV/Vis and Fluorescence Measurements Concerning the Formation of PDI/FLG Hybrids in Solution

For PDIs **1**–**3**, the best conditions for successful exfoliation and functionalization of turbostratic graphite were found to be those where the PDI had a concentration of $10^{-5}$ M ($2.65 \cdot 10^{-2}$ mg/mL) and the graphite 1.2 mg/mL, which translates in a PDI/FLG mass ratio of 0.0221.

Comparing the absorption spectra of the PDIs alone after stirring (Table 1, top) and those of the PDI/FLG hybrids in solution (Table 1, bottom), a dramatic red-shift ($\approx$ 30 nm) of the absorption bands accompanied by a strong quenching of the fluorescence of PDIs was observed in all cases. Derivatives **1** and **2** appear to interact most strongly with the carbon exfoliated material as their fluorescence quenching is accompanied also by a slight bathochromic shift of the fluorescence peaks of the PDI and their fluorescence quantum yields in presence of graphite were recorded to be approximately 100 times lower than those documented for PDI **3**.

The fluorescence quenching of the PDIs is most probably the result of a FRET process between the PDI core of the surfactants and the exfoliated graphene material,[1] while the red-shifts in both absorption and emission bands indicate that electronic transfers are taking place as well upon adhesion of the PDI moieties on top of the exfoliated graphitic material in aqueous solution.[2]

The absorption and emission features of **1**–**3** in presence and absence of graphite are summarized in Table 1.

| | $\lambda_1$ (nm) | $\lambda_2$ (nm) | $A_1$ | $A_2$ | $A_2/A_1$ | $F_1$ | $F_2$ | $\Phi_{\lambda_1}$[a] |
|---|---|---|---|---|---|---|---|---|
| **1** | 498 | 537 | 0.28 | 0.15 | 0.53 | 545 | 587 | 0.04 |
| **2** | 498 | 533 | 0.33 | 0.35 | 1.06 | 545 | 588 | 0.39 |
| **3** | 498 | 534 | 0.38 | 0.4 | 1.05 | 545 | 588 | 0.45 |

| | $\lambda_1$ (nm) | $\lambda_2$ (nm) | $A_1$ | $A_2$ | $A_2/A_1$ | $F_1$ | $F_2$ | $I_F/I_0$ | $\Phi_{\lambda_1}$[a] |
|---|---|---|---|---|---|---|---|---|---|
| **1**-graphene | -[b] | 560 | - | 0.63 | - | 550 | 597[d] | $5.5 \cdot 10^{-3}$ | $9.1 \cdot 10^{-5}$ |
| **2**-graphene | 500 | 560 | 0.46 | 0.44 | 0.96 | 551 | 597 | $4 \cdot 10^{-4}$ | $9.7 \cdot 10^{-5}$ |
| **3**-graphene | 500 | 560[c] | 0.53 | 0.50 | 0.94 | 545 | 588 | $1.5 \cdot 10^{-2}$ | $4.8 \cdot 10^{-3}$ |

**Table S1.** Top: Absorption and emission features of PDI **1**–**3** after magnetic stirring ($c_{PDI} = 10^{-5}$ M). Bottom: Absorption and emission features of PDI/FLG hybrids in solution after magnetic stirring and centrifugation ($c_{PDI} = 10^{-5}$ M, $c_{G,i} = 1.2$ mg/mL). a) The fluorescence quantum yield ($\Phi$) is calculated taking fluorescein in NaOH 0.1 M as reference (*Photochem. Photobiol.* **2002**, *75*, 327); b) peak not clearly discernible; c) a shoulder peak ($\lambda \approx 535$ nm) seems to be present as well; d) a broad excimer band at 660-670 nm is visible in the spectrum.

## 3. Investigation of the Stability of the PDI/FLG Hybrids Against Washing

In order to acquaint for the stability of the PDI/FLG hybrids against washing in the solid state, the wafers prepared for the Raman characterization were washed by spin coating with water (1 mL) and with methanol (1 mL). After each washing step Raman characterization was performed ($\lambda_{exc}$ = 532 nm). As depicted Figure S8, the PDI/FLG hybrids resulted to be very stable against washing as the Raman features proper to both PDI and FLG components are always present in the spectra.

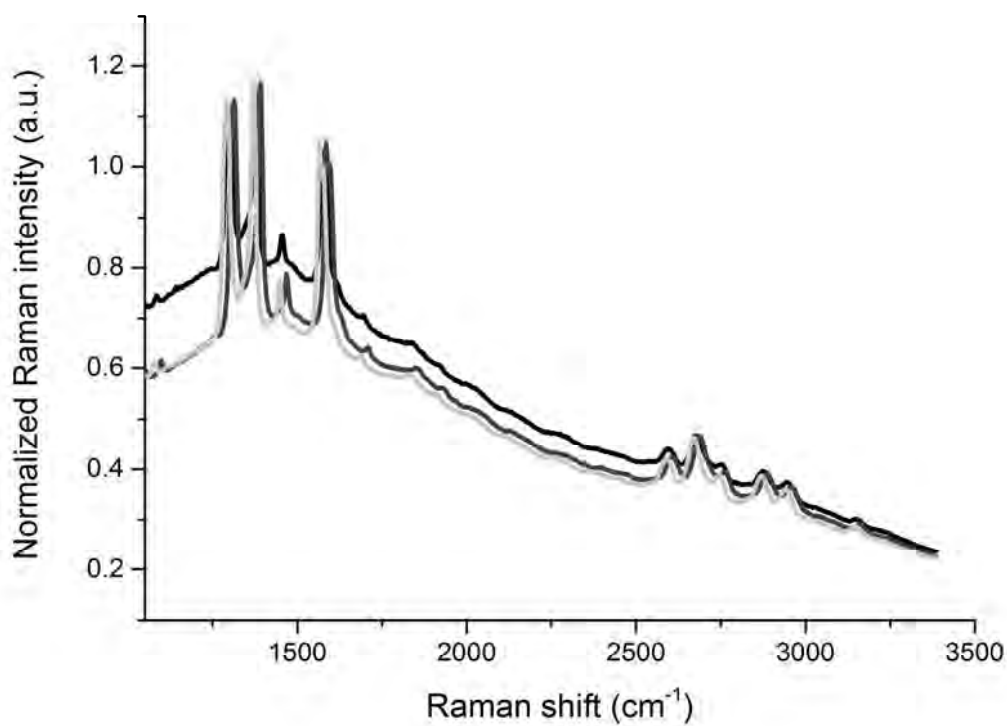

**Figure S8.** Mean Raman spectra for **1**/FLG hybrids after drop-casting of 1 μL of the freshly centrifuged dispersion on top of a Si/SiO$_2$ wafer and drying (light grey trace); after washing with 1 mL of pure water (dark grey trace) and after washing with 1 mL of pure water and 1 mL of ethanol (black trace). The Raman spectra presented are the respective mean spectra of 1680 recorded single-point spectra and they were normalized for the intensity of the G-band; $\lambda_{exc}$ = 532 nm.

## 4. Time-Dependent Analysis and Determination of the Concentration of FLG in Solution

In order to further improve the non-covalent functionalization of the FLGs, time dependent dispersion experiments were performed. Namely, a solution of **1** (c = $10^{-5}$ M) containing turbostratic graphite (c = 1.2 mg/mL) was stirred for a certain time (6, 24, 30 and 48 h) and then immediately centrifuged avoiding the sedimentation step. The dispersions were characterized by means of absorption spectroscopy before and after centrifugation in order to determine the dispersion efficiency (DE) of the PDI surfactant and the graphene concentration in solution ($c_{G, f}$). The DE value (%) was calculated by dividing the absorbance collected at 660 nm after centrifugation for that measured before. The final graphene concentration was determined by dividing the absorbance at 660 nm after centrifugation for the extinction coefficient for graphene ($\alpha_{660}$). For $\alpha_{660}$, a value of 1448 mL·mg$^{-1}$·m$^{-1}$ was used, calculated as an average between those reported by Lotya et al.[3] and Zhang et al.[4] As depicted in Figure S9, upon exfoliation and non-covalent functionalization with PDI **1**, both DE and $c_{G, f}$ increases till ≈ 31 h of stirring and then slightly decreases.

However, for practical reasons and due to the fact that the increase in DE between 24 h and 31 h is no more proportional to the increase of stirring time, it has been decided to employ a stirring time of 24 h followed by centrifugation to prepare the PDI-FLG dispersions needed for the iterative dip-coating experiments.

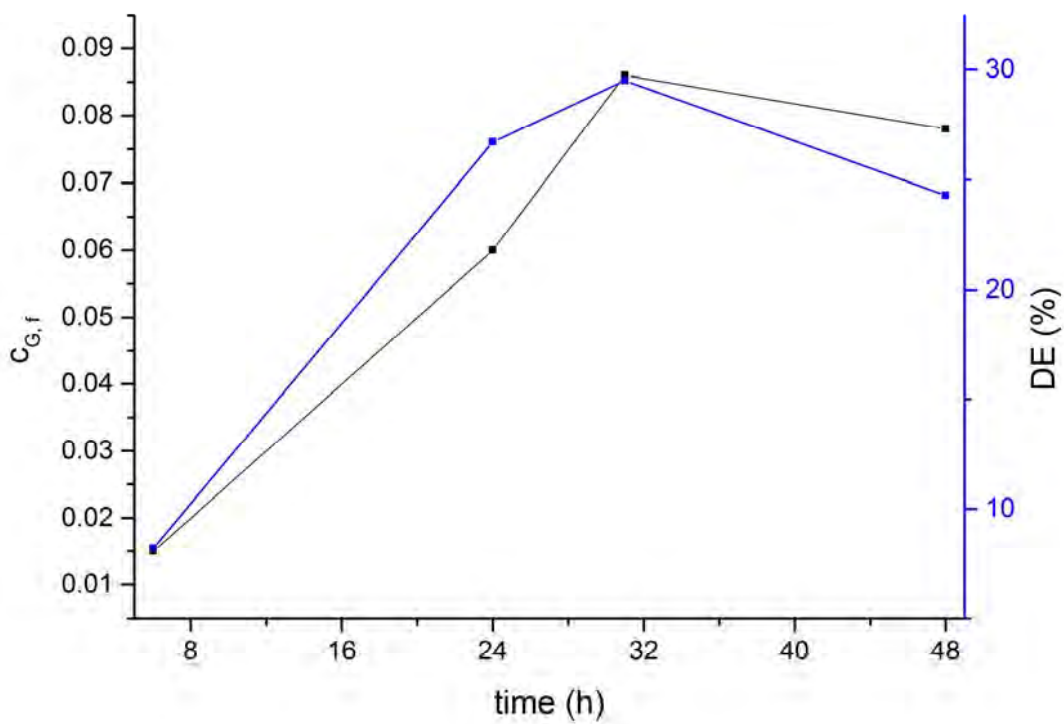

**Figure S9.** Relation between the dispersion efficiency (DE, blue line) of PDI **1** and the final concentration of the FLG sheets in solution after centrifugation ($c_{G, f}$, black line) by changing the stirring time. In all experiments the concentration of PDI **1** was $10^{-5}$ M, while the initial concentration of turbostratic graphite 1.2 mg/mL.

The DE and $c_{G,f}$ values obtained after 24 h of stirring turbostratic graphite in presence of PDIs **1 – 3** are reported in Table 2.

|  | DE (%) | $c_{G,f}$ (mg/mL) |
|---|---|---|
| **1**/FLG | 26.7 | 0.06 |
| **2**/FLG | 12.8 | 0.03 |
| **3**/FLG | 5.2 | 0.01 |

**Table S2.** DE and $c_{G,f}$ values for PDIs **1 – 3** after centrifugation ($c_{PDI} = 10^{-5}$ M; $c_G$ = 1.2 mg/mL; stirring time 24 h)

Altogether, it can be summarized that the dispersion ability for the cationic PDIs employed in this study follows the order **1 > 2 > 3**. As recalled from Table 1, PDI **1** showed the highest tendency towards aggregation in pure aqueous environments and its aggregates were stable against stirring. However, this was not the case for **2** and **3**. Moreover, Schönamsgruber *et al.*[5] explained that an elongation in the spacer unit chain resulted in better solubility of the PDI and thus less aggregation. Our results show that the aggregation of π-based surfactants is a central key point to allow the successful dispersion and functionalization of graphene structures in aqueous solution and are in perfect agreement with our recent reports concerning the dispersion and non-covalent functionalization of CNTs.[6] Additionally, preliminary tests performed in buffered and non-buffered solutions has outlined that derivative **1** could be able to efficiently disperse CNTs as well but that its performances were strictly pH and ionic strength dependent.

## 5. Characterization of the ZnO-NP

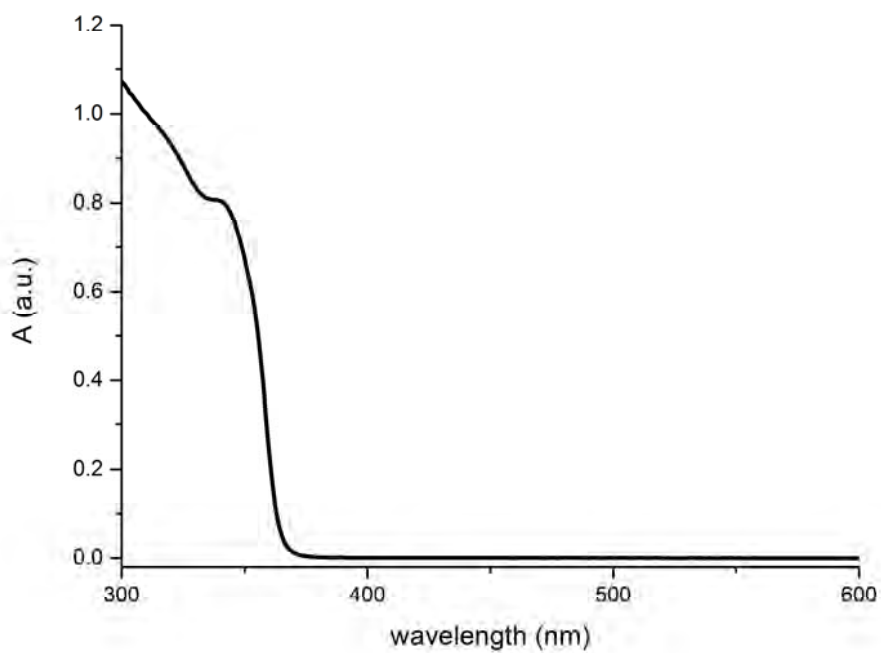

**Figure S10.** UV/Vis absorption spectrum of ZnO-NP in ethanol (c = 0.05 mg/mL).

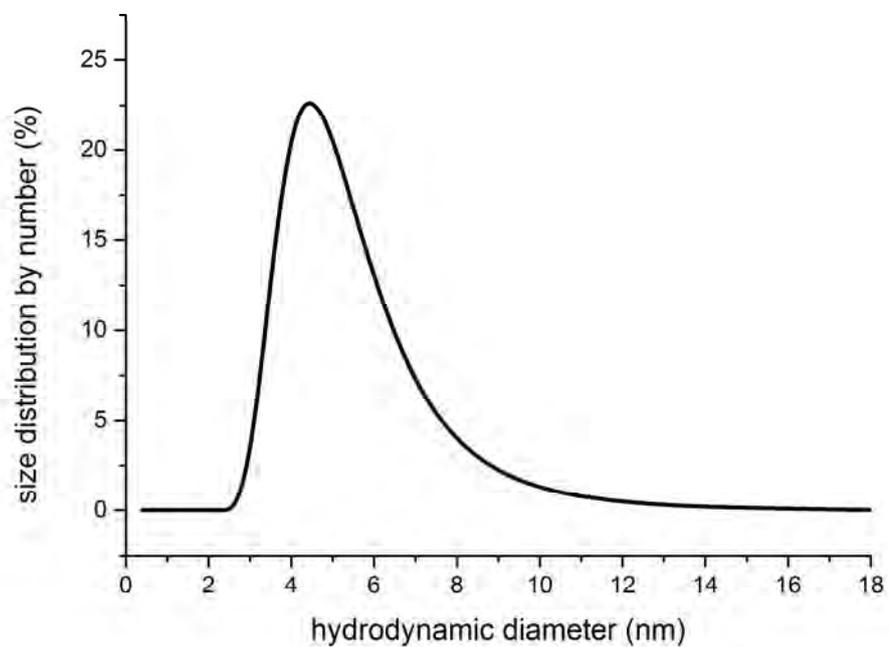

**Figure S11.** Number size distribution for ZnO-NP in ethanol (from DLS measurement).

## 6. Characterization of the [ZnO-NP/PDI/FLG]$_n$ Nano-Structures

### 6.1 UV/Vis Spectroscopy

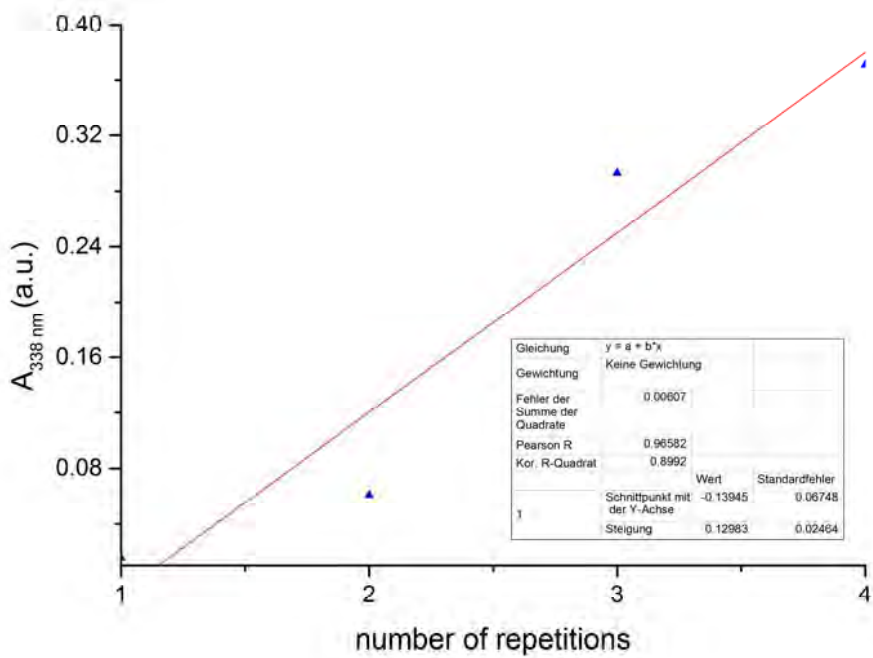

**Figure S12.** Experimental values and linear fitting of the UV/Vis absorption of ZnO-NP/**1**/FLG assemblies at 338 nm, against the number of repetitions.

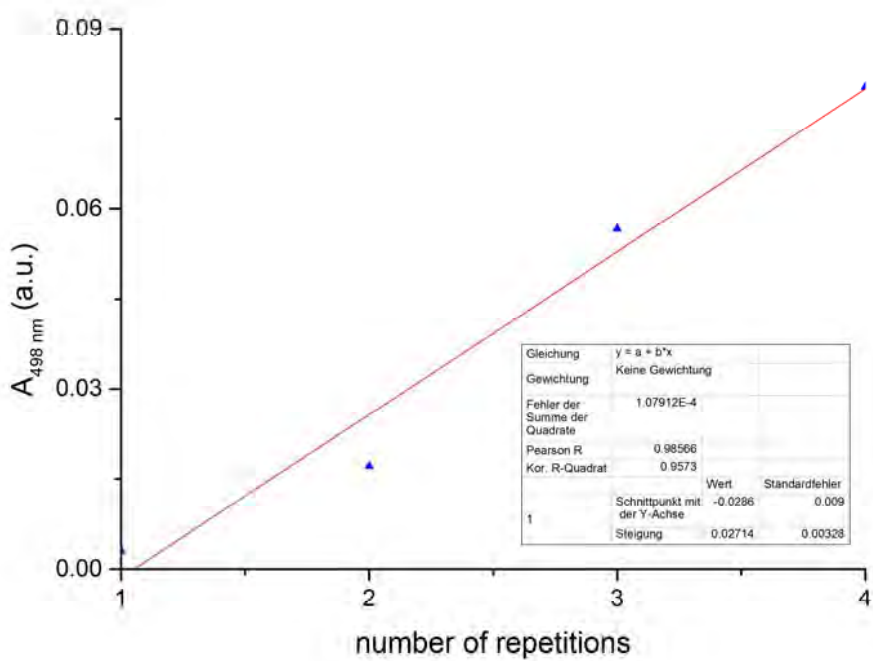

**Figure S13.** Experimental values and linear fitting for the UV/Vis absorption of ZnO-NP/**1**FLG assemblies at 498 nm, against the number of repetitions.

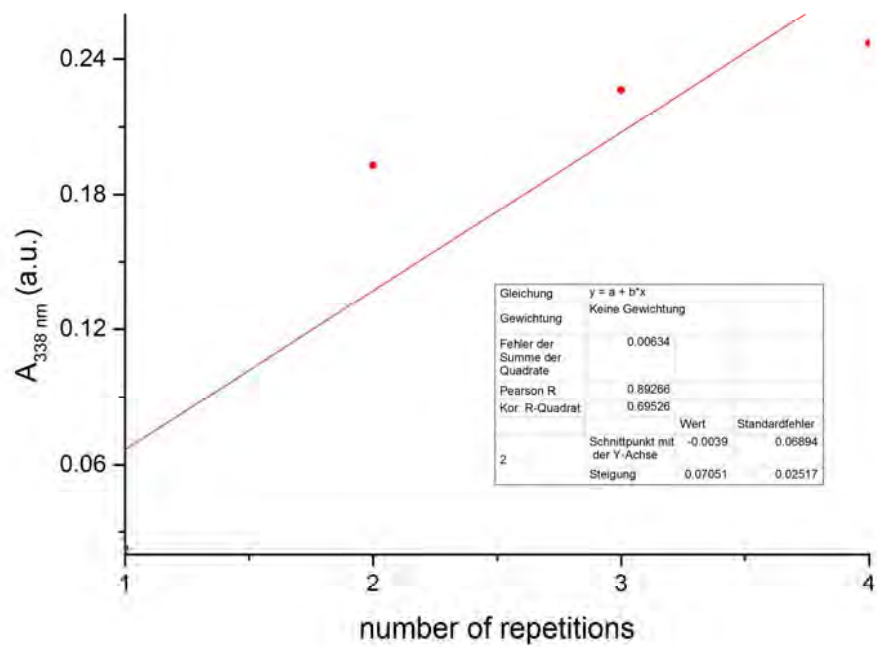

**Figure S14.** Experimental values and linear fitting for the UV/Vis absorption of ZnO-NP/**2**/FLG assemblies at 338 nm, against the number of repetitions.

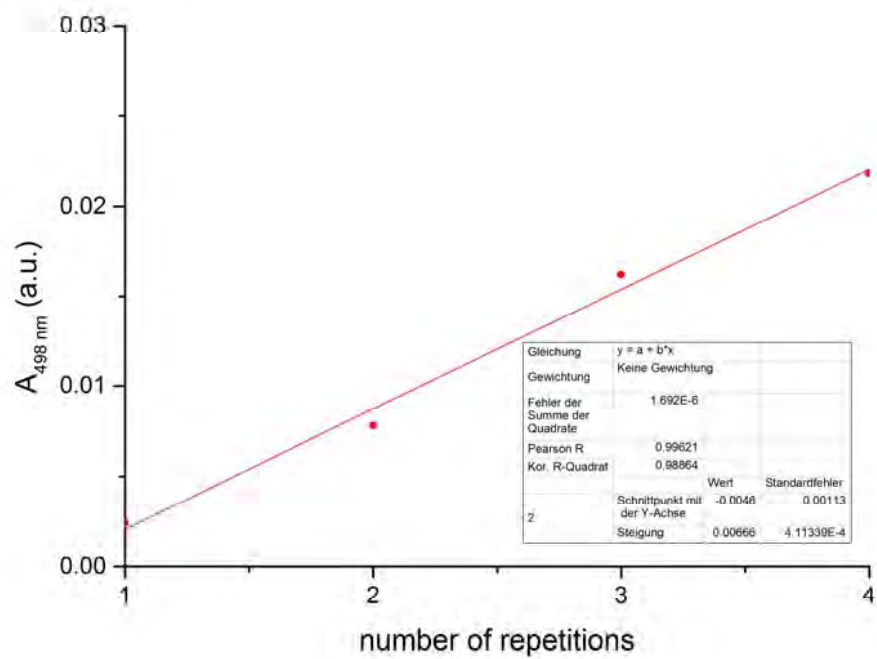

**Figure S15.** Experimental values and linear fitting for the UV/Vis absorption of ZnO-NP/**2**/FLG assemblies at 498 nm, against the number of repetitions.

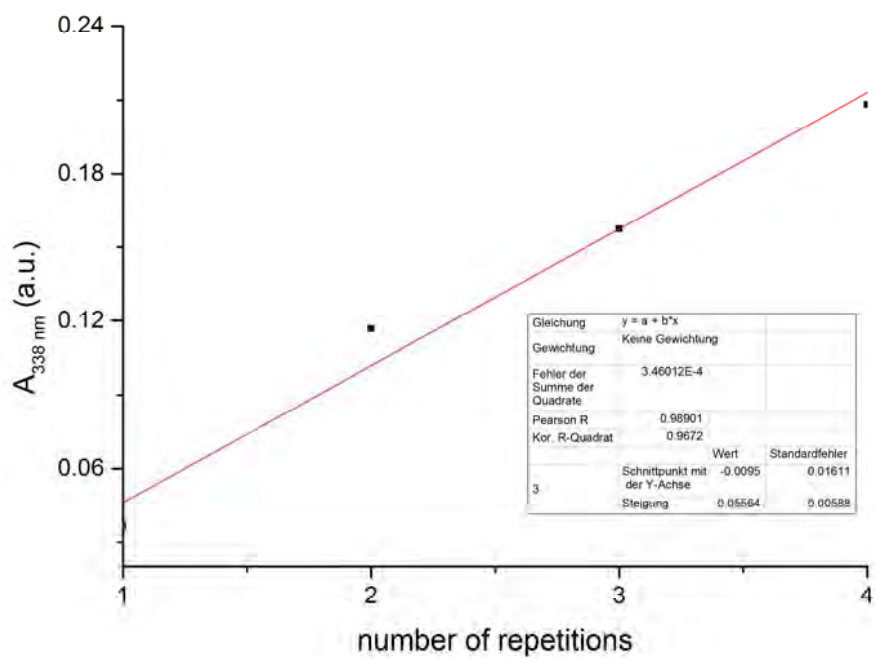

**Figure S16.** Experimental values and linear fitting for the UV/Vis absorption of ZnO-NP/**3**/FLG assemblies at 338 nm, against the number of repetitions.

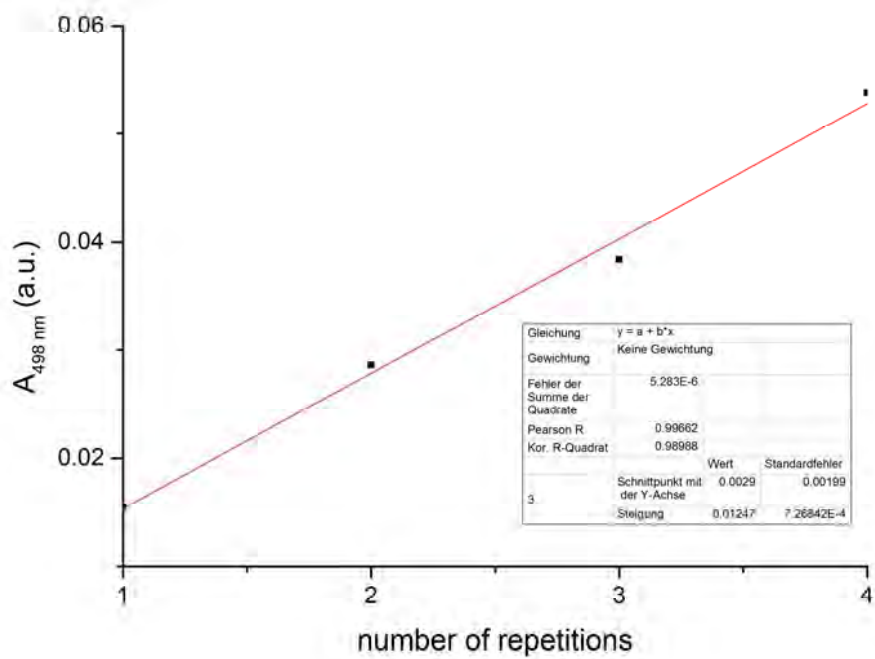

**Figure S17.** Experimental values and linear fitting for the UV/Vis absorption of ZnO-NP/**3**/FLG assemblies at 498 nm, against the number of repetitions.

## 6.2 AFM

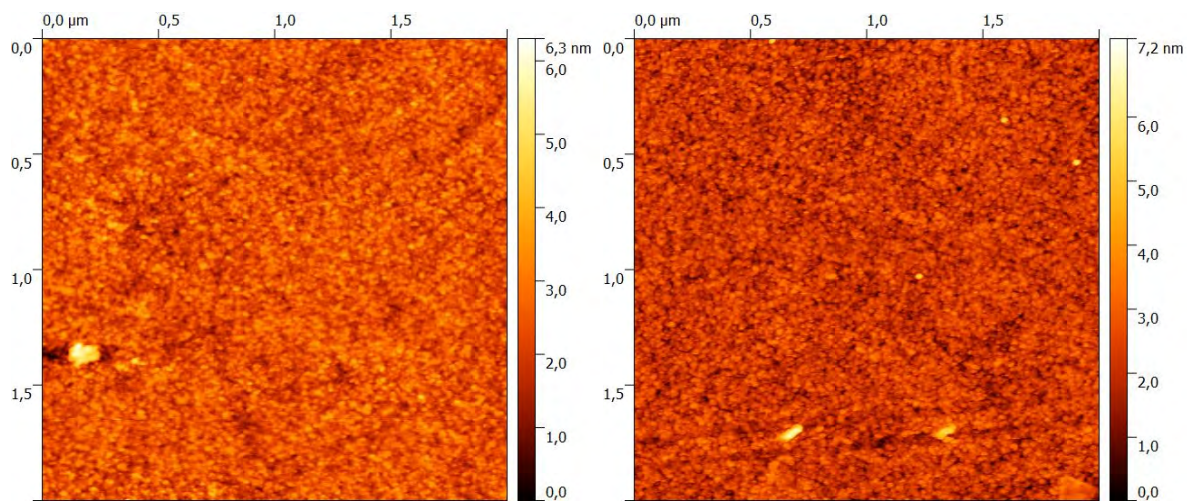

(left) SiO$_2$ pristine, RMS= 0.170 nm; (right) SiO$_2$ etched RMS=0.212 nm.

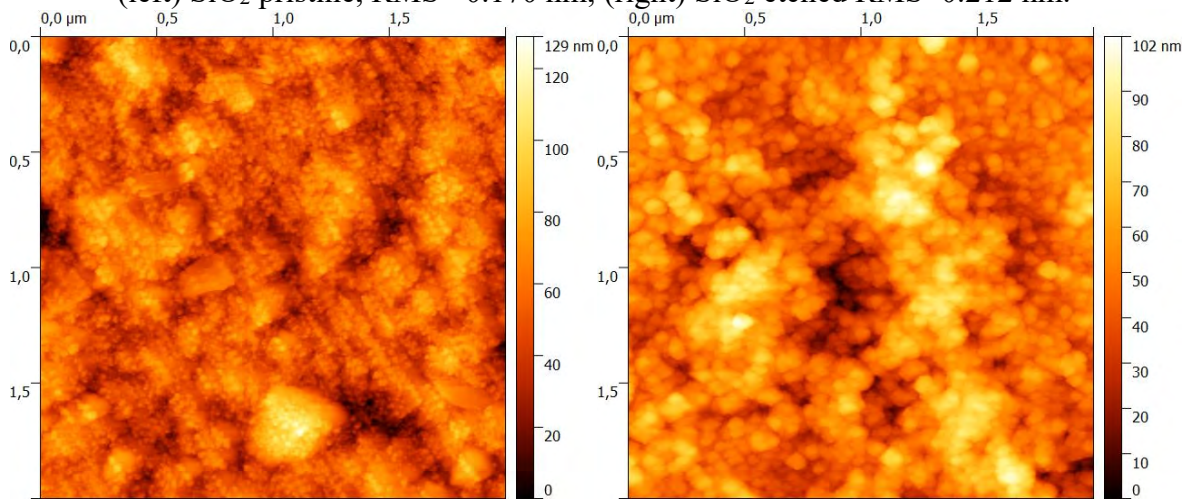

(left) [ZnO-NP/**1**/FLG]$_4$, RMS =6.10 nm; (right) [ZnO-NP/**1**]$_4$, RMS= 10.1 nm.

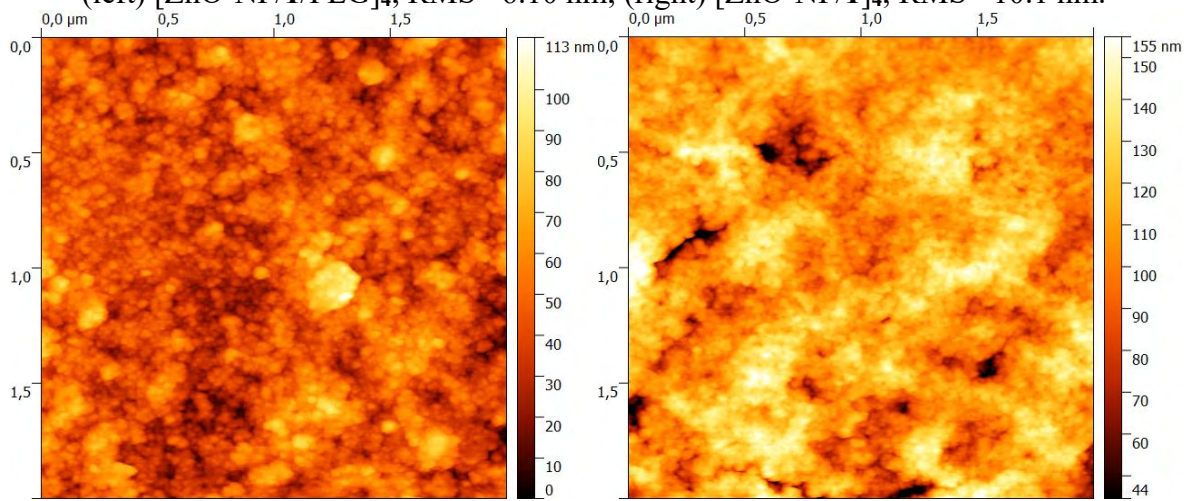

(left) [ZnO-NP/**2**/FLG]$_4$, RMS =6.93 nm; (right) [ZnO-NP/**2**]$_4$, RMS= 11.1 nm.

**Figure S18.** AFM pictures with relative root mean square (RMS) roughness values.

6.3 Electron Microscopy

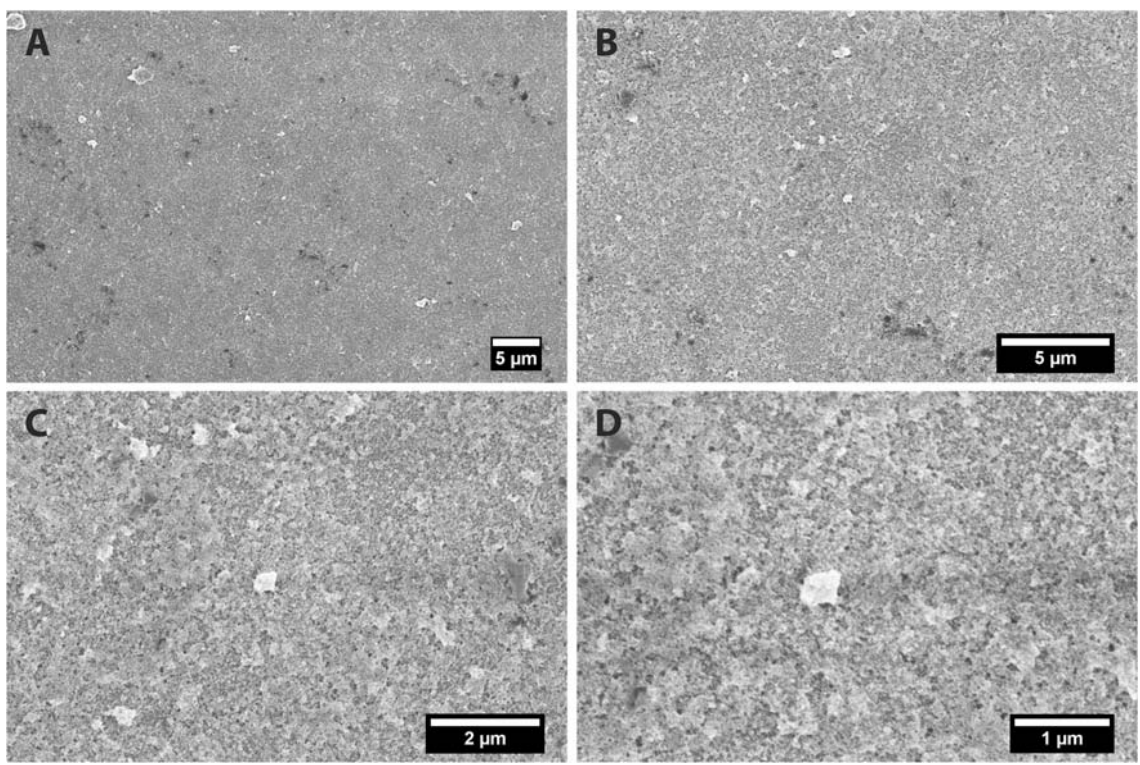

**Figure S19.** Top-view on the [ZnO-NP/**1**/FLG]$_4$ hybrid assembly acquired with scanning electron microscopy (SEM) showing different magnifications: imaging at 5 kV, secondary electrons. Field width 59.2 μm (A), 25.9 μm (B), 10.4 μm (C), 5.92 μm (D).

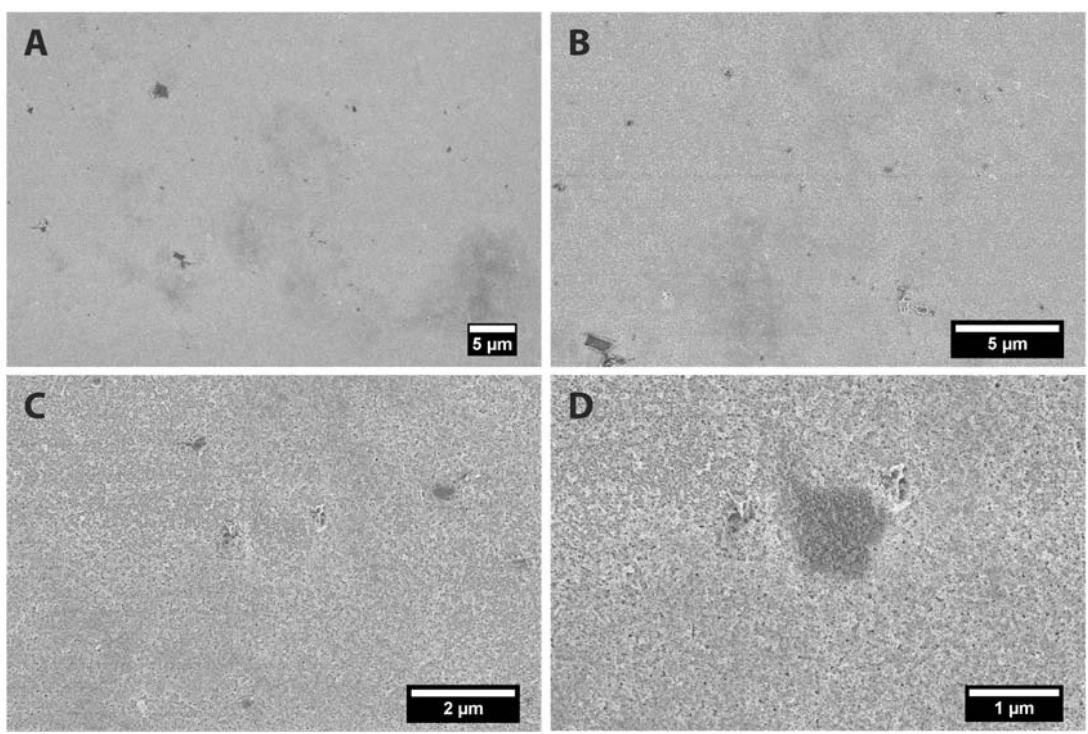

**Figure S20.** Top-view on the [ZnO-NP/**2**/FLG]$_4$ hybrid assembly acquired with scanning electron microscopy (SEM) showing different magnifications: imaging at 2 kV, secondary electrons. Field width 59.2 μm (A), 25.9 μm (B), 10.4 μm (C), 5.92 μm (D).

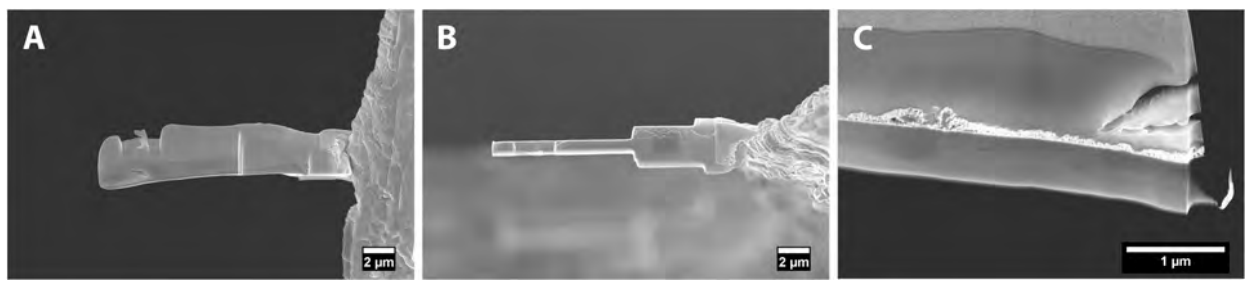

**Figure S21.** Lamella during preparation imaged with SE: Side view (A), top view (B), close-up of the thin film (C).

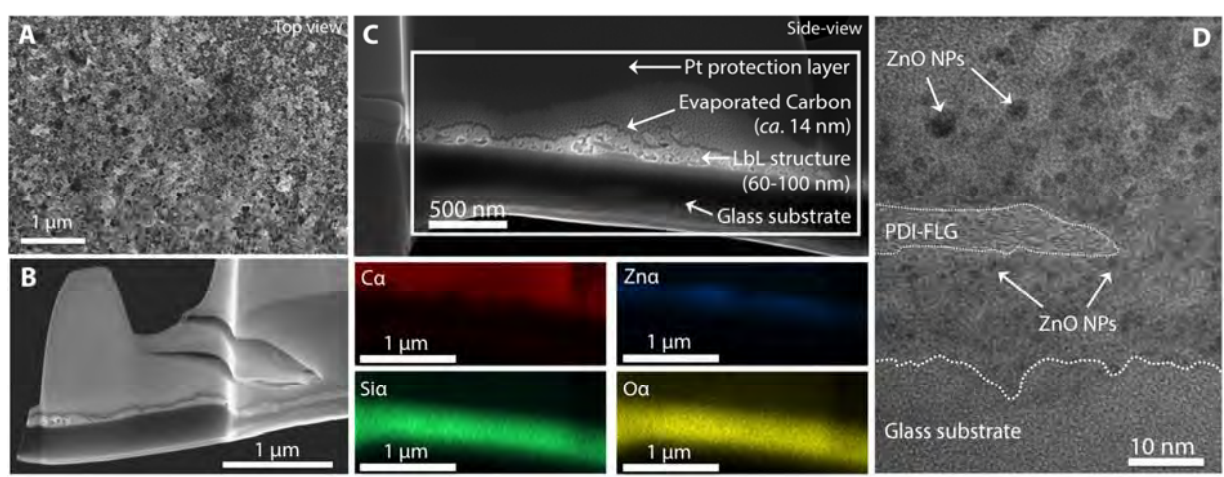

**Figure S22**. Additional data for the investigation of the lamella: A) Low-magnification SEM image showing the top view of the film. B) Final cross-section slice prepared by means of focused ion beam SEM. C) Higher magnification and EDX mappings of the cross-section, highlighting the presence of the ZnO-NP and the PDI/FLG in the elemental chemical analysis. D) HRTEM image of the lamella after post thinning of the sample, revealing the intimate contact

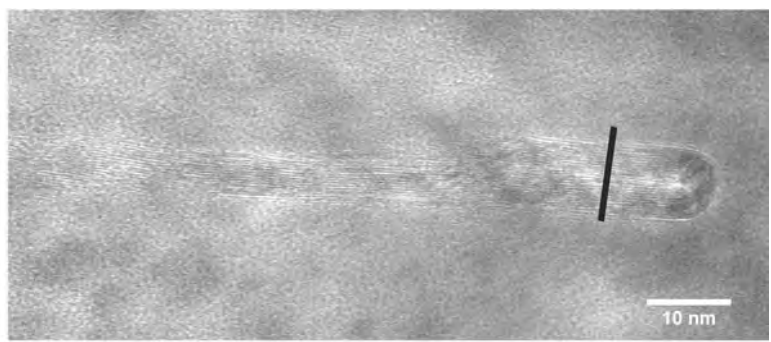
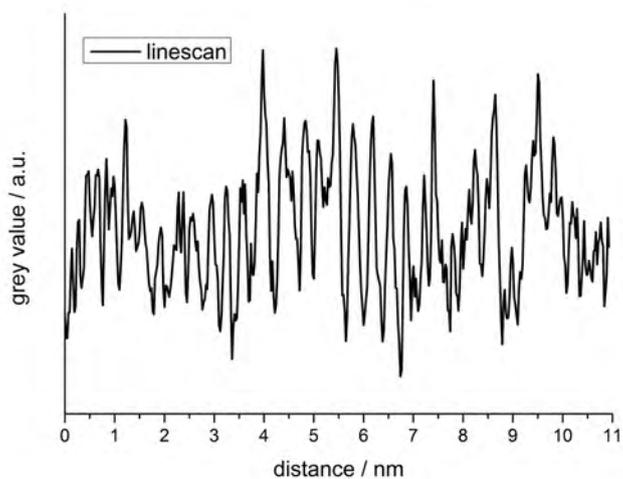

**Figure S23.** HRTEM image and indicated linescan over a folded FLG flake showing the graphitic nature of the carbon material and interlayer spacing.

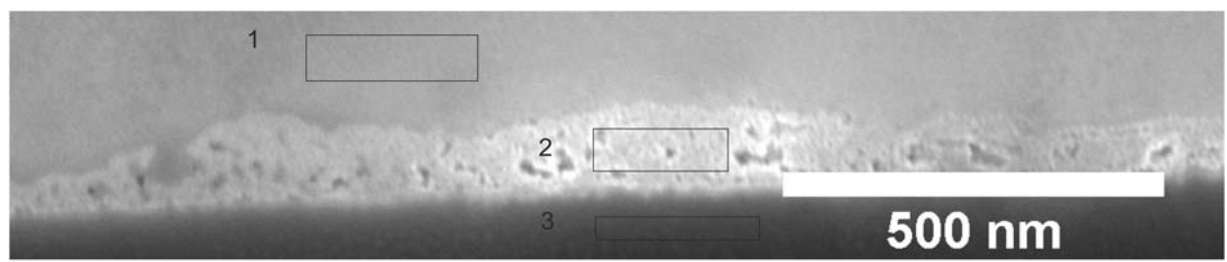
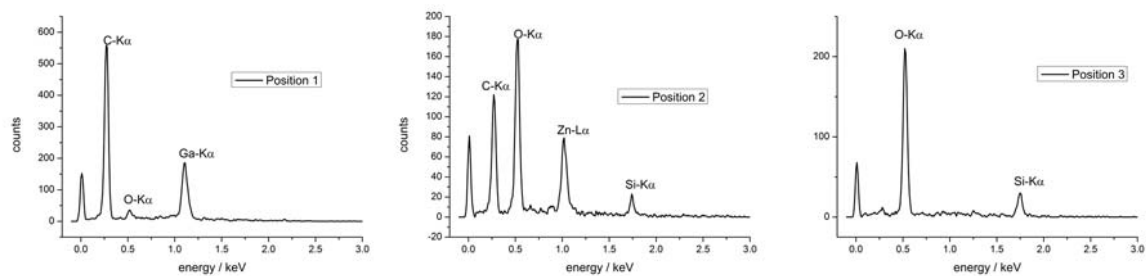

**Figure S24.** EDX point spectra analysis of the cross-section from indicated positions on the sample. Above film (position 1), in the [ZnO-NP/PDI/FLG]$_4$ film (position 2; presence of PDI/FLG and ZnO-NP), in the glass substrate (position 3).

# 7. Rational Explanation for the Suitable PDI/FLG Mass Ratio

In section 3.1 we outlined that the best PDI/FLG ratio to determine successful dispersion and non-covalent functionalization of graphene in solution resulted to be about 0.022 ($c_{G,i}$ = 1.2 mg/mL with a $c_{PDI}$ = $10^{-5}$ M).

Here we show, through simple calculations, that by comparison between the specific surface area of the graphite ($A_G$) and the theoretical area offered by the PDI-cores (PTCDI units) of the surfactant molecules ($A_{PTCDI}$) a correlation can be found to determine the conditions for optimal exfoliation/functionalization.

As reported by Kozhemyakina et al.[7] the specific surface area of turbostratic graphite is 16.3 m$^2$/g. Ludwig et al.[8] calculated by means of STM investigations the lattice parameter for PTCDI on highly oriented pyrolytic graphite (HOPG). The area of the unit cell was determined to be 243.7 Å$^2$ and two molecules are contained in the unit cell. Therefore, we can reasonably define the area of one PTCDI unit as 121.85 Å$^2$ ≈ 1.22 nm$^2$.

If we would like to disperse 12 mg of turbostratic graphite in 10 mL of water ($c_G$ = 1.2 mg/mL), we would obtain a value for $A_G$ of about 1.96·10$^{17}$ nm$^2$. In Table 3, we have summarized the values for $A_{PTCDI}$ calculated, starting from three different solutions of PDI with increasing concentration, $c_{PDI}$ = $10^{-6}$ M – $10^{-4}$ M:

|  | $A_{PTCDI}$ (nm$^2$) |
|---|---|
| $10^{-6}$ M | 7.35·10$^{15}$ |
| $10^{-5}$ M | 7.35·10$^{16}$ |
| $10^{-4}$ M | 7.35·10$^{17}$ |

**Table S3.** $A_{PTCDI}$ values for three different concentrations of PDI.

If we compare now the three values for $A_{PTCDI}$ with that determined for $A_G$ (1.96·10$^{17}$ nm$^2$), we can see in one case ($c_{PDI}$ = $10^{-5}$ M) $A_{PTCDI}$ ≈ $A_G$, while in the two other cases the value for $A_{PTCDI}$ is either too high or too low with reference to $A_G$.

From our experimental results we can therefore conclude that a successful exfoliation/functionalization of graphite with PDIs can only occur when the π-surface available from graphite is comparable to that offered from the dye molecules. Instead, a too high or too low surface ratio ($A_{PTCDI}$/$A_G$) results experimentally in the presence of too much free PDI molecules in solution or in the precipitation of all carbon material, respectively.

We are conscious that our calculations are quite simplified as they do not take into account aggregation phenomena in solution, which play indeed a key role in determining the true efficiency of the PDI dyes with respect to successful exfoliation/functionalization of graphene. Nevertheless, they offer a first qualitative indication to rationally determine suitable π-surfactant/graphite ratios to allow the dispersion of graphite material in solution.